\documentclass[aps,amsmath,twocolumn,amssymb,floatfix,showpacs,superscriptaddress,nofootinbib,longbibliography]{revtex4-1}
\usepackage{mathtools}
\usepackage{braket}
\usepackage[dvipsnames]{xcolor}
\usepackage{float}
\usepackage{subfigure}
\usepackage{dsfont}
\usepackage{MnSymbol}
\usepackage{braket}
\usepackage[dvipsnames]{xcolor}
\usepackage{float}
\usepackage{subfigure}
\usepackage{tikz}
\usepackage[colorlinks=true,linktoc=page,linkcolor=red,citecolor=violet,urlcolor=purple]{hyperref}
\usepackage{multirow}
\newcommand{\vect}[1]{\boldsymbol{\mathrm{#1}}}

\mathchardef\mhyphen="2D 

\newcommand{\ie}{{i.e.,\,\,}}
\newcommand{\eg}{{e.g.,~}}
\newcommand{\ua}{{\uparrow }}
\newcommand{\da}{{\downarrow }}
\newcommand{\eps}{{\epsilon}}

\newcommand\bea{\begin{eqnarray}}
	\newcommand\eea{\end{eqnarray}}
\newcommand\beq{\begin{equation}}  
	\newcommand\eeq{\end{equation}}

\newcommand{\non}{\nonumber}  
\newcommand{\dis}{\displaystyle}
\newcommand{\Schr}{Schr\"{o}dinger \,}

\newcommand{\ch}{\mathcal{S}}%
\newcommand{\Wz}{\mathcal{W}_0} 
\newcommand{\Wp}{\mathcal{W}_{\pi}} 

\newcommand{\mc}{\mathcal}

\newcommand{\A}{\alpha}
\newcommand{\jj}{\mathrm{JJ}}
 
\newcommand{\tbf}{\textbf}
\usepackage[normalem]{ulem}
\definecolor{lime}{HTML}{A6CE39}
\usepackage{sidecap,tikz}
\DeclareRobustCommand{\orcidicon}{\hspace{-1.0mm}
	\begin{tikzpicture}
		\draw[lime, fill=lime] (0.0,0.0) 
		circle [radius=0.15] 
		node[white] {{\fontfamily{qag}\selectfont \tiny \,ID}};
		\draw[white, fill=white] (-0.0525,0.095) 
		circle [radius=0.007];
	\end{tikzpicture}
	\hspace{-3.0mm}
}
\foreach \x in {A, ..., Z}{\expandafter\xdef\csname orcid\x\endcsname{\noexpand\href{https://orcid.org/\csname orcidauthor\x\endcsname}{\noexpand\orcidicon}}
}

\AtBeginDocument{%
	\newwrite\bibnotes
	\def\bibnotesext{Notes.bib}
	\immediate\openout\bibnotes=\jobname\bibnotesext
	\immediate\write\bibnotes{@CONTROL{REVTEX41Control}}
	\immediate\write\bibnotes{@CONTROL{%
			apsrev41Control,author="08",editor="1",pages="1",title="1",year="1"}}
	\if@filesw
	\immediate\write\@auxout{\string\citation{apsrev41Control}}%
	\fi
}%
\begin{document}

	
	\title{Josephson current signature of Floquet Majorana  and topological accidental zero modes in altermagnet heterostructures}  
	
	\author{Amartya Pal\orcidA}
	\affiliation{Institute of Physics, Sachivalaya Marg, Bhubaneswar-751005, India}
	\affiliation{Homi Bhabha National Institute, Training School Complex, Anushakti Nagar, Mumbai 400094, India}

	\author{Debashish Mondal\orcidD{}}
	\affiliation{Institute of Physics, Sachivalaya Marg, Bhubaneswar-751005, India}
	\affiliation{Homi Bhabha National Institute, Training School Complex, Anushakti Nagar, Mumbai 400094, India}
	
	\author{Tanay Nag\orcidB{}}
	\email{tanay.nag@hyderabad.bits-pilani.ac.in}
	\affiliation{Department of Physics, BITS Pilani-Hydrabad Campus, Telangana 500078, India}
	
	\author{Arijit Saha\orcidC{}}
	\email{arijit@iopb.res.in}
	\affiliation{Institute of Physics, Sachivalaya Marg, Bhubaneswar-751005, India}
	\affiliation{Homi Bhabha National Institute, Training School Complex, Anushakti Nagar, Mumbai 400094, India}

\begin{abstract}
We theoretically investigate the generation and Josephson current signatures of Floquet Majorana end modes (FMEMs) in a periodically driven altermagnet (AM) heterostructure. Considering a one-dimensional (1D) Rashba nanowire (RNW) proximitized to a regular $s$-wave superconductor and a $d$-wave AM, we generate both $0$- and $\pi$-FMEMs by driving the nontopological phase of the static system. While the static counterpart hosts both topological Majorana zero modes (MZMs) and nontopological accidental zero modes (AZMs), the drive can gap out the static AZMs and generate robust $\pi$-FMEMs, termed as topological AZMs (TAZMs). We topologically characterize the emergent FMEMs via dynamical winding numbers exploiting chiral symmetry of the system. Moreover, we consider a periodically driven Josephson junction comprising of RNW/AM-based 1D topological superconduting setup.  We identify the signature of MZMs and FMEMs utilizing $4\pi$-periodic Josephson effect, distinguishing them from trivial AZMs exhibiting $2\pi$-periodicty, in both static and driven platforms. This Josephson current signal due to Majorana modes survives even in presence of finite disorder. Our work establishes a route to realize and identify FMEMs in AM-based platforms through Floquet engineering and Josephson current response.
\end{abstract}
%
%
	\maketitle
%
\textit{\textcolor{blue}{Introduction.---}} The hunt for of realizing topological superconductors (TSCs) hosting Majorana zero modes (MZMs) has driven intense research interest for decades. The nonAbelian braiding statistics of MZMs enable them as a promising candidate for decoherence-free topological quantum computation~\cite{Kitaev_2001,Leijnse_2012,Alicea2011_NatPhys,Alicea_2012,LutchynPRL2010,Beenakker2013search,TewariPRL2012,Kitaev2003_annals_of_phys,NayakRMP2008,Qi2011_MZM,Aguado2017,Yazdani_hunt_majorana,PritamPRB2023,ZhongboYan2019,Wang2022_PRL,Minakshi2024_APL}. MZMs, the charge-neutral zero-energy quasiparticle excitations in TSCs, were theoretically proposed in 1D spinless $p$-wave superconductor (SC) by Kitaev~\cite{Kitaev_2001}, and later proposed to be engineered in Rashba nanowires (RNWs) with strong spin-orbit coupling (SOC), placed in close proximity to a $s$-wave SC under an external Zeeman field~\cite{Leijnse_2012,Alicea2011_NatPhys,Alicea_2012,LutchynPRL2010}.  Owing to resonant Andreev reflection, a pair of MZMs exhibit zero bias conductance peak (ZBCP) at zero temperature, quantized at $2e^2/h$ \cite{LawPRL2009,Mondal2025PRBL,Rokhinson2012,Finck2013,Albrecht2016,Das2012_NatPhys,Mourik2012Science,Deng2016,JunSciAdv2017}. Experimental detection of such ZBCP via differential conductance ($dI/dV$) measurement is often considered as indirect signature of MZMs~\cite{Das2012_NatPhys,Mourik2012Science,Deng2016,JunSciAdv2017}. However, various alternative phenomena \eg Andreev bound states~\cite{KellsPRB2012,Lee2014,Ricco_ABS_PRB2019}, Kondo resonance~\cite{GoldhaberNature1998,Cronenwett1998Kondo}, disorder~\cite{BagretsPRL2012} etc., can also produce a ZBCP, leading to dispute in MZMs detection. In sharp contrast, the $4\pi$-periodic Josephson effect serves as a promising signature of MZMs~\cite{Alicea2011_NatPhys,Alicea_2012,Beenakker2013search,Fu2009, Law2011_PRB,SDSarma2019_JJ,LutchynPRL2010,Trauzettel2014_PRL,Braggio2016_FP,Cayao2015,Cayao2017,Cayao2019} and cannot be replicated in other physical systems. When two weakly coupled TSCs with a superconducting phase difference, $\phi$, form a Josephson junction (JJ), the Josephson current (JC) exhibits $4\pi$ periodicity due to tunneling of MZMs across the junction unlike the usual $2\pi$-periodicity in conventional JJs, where only Cooper pairs contribute~\cite{Alicea2011_NatPhys,Beenakker2013search}. This $4\pi$-periodicity originates from the global conservation of ground state fermionic parity~\cite{Alicea2011_NatPhys,Fu2009}.

In current literature, altermagnets (AMs) are proposed to be a new class of antiferromagnets with even parity collinear-compensated magnetic order where two opposite spin sublattices are connected through rotation rather than only translation \eg $C_{4}$ rotation in $d$-wave AMs~\cite{Smejkal_PRX_1,Smejkal_PRX_2,BhowalPRX2024,Bai_PRL_2023,Mazin_PRBL_2023,XZhou2024_PRL,Lin2025}. Interestingly, regardless of the net zero magnetization, AMs break time-reversal symmetry (TRS) and host unique spin-split band structures due to anisotropic momentum-dependent exchange interactions as observed in $\mathrm{RuO_2, MnTe, CrSb, MnF_2}$ etc.~\cite{Smejkal_PRX_1,Smejkal_PRX_2,BhowalPRX2024,Bai_PRL_2023,Lee2024PRL,Jiang2025}. 
Very recently, generation of MZMs and nontopological accidental zero modes (AZMs) using AMs has been proposed by replacing the external Zeeman field in 1D RNW system~\cite{Ghorashi2024PRL,Mondal2025PRBL,Li_PRBL_2023,Li_PRBL_2024,Zhu2023,Li2024,Zhang2024,Yin2025PRB,Fukaya2025arXiv,Maeda2025arXiv,Chatterjee2025arXiv,Li2025_arXiv}. The net zero magnetization is advantageous for sustaining the proximity induced superconductivity. In recent times, various phenomena in AM-based JJ is also investigated in literature~\cite{Lu2025PRL,Fukaya2025PRB_JJ,Trauzettel2024arXiv,Ouassou2023PRL}. However identifying MZMs and AZMs in the JJ setup based on AM-SC heterostructure remains to be addressed. 

On the other hand, one promising approach to engineer Majorana modes is via Floquet engineering where a time periodic drive generates topologically nontrivial states in quantum materials~\cite{Oka2009_FloquetGraphene,Mikami16,lindner11floquet,Rudner2013,Usaj2014,Asboth2014,Perez2014,Eckardt_2015,Oka2019,Yao2017,Rudner2020,Nag2021anomalous,benito14,Ghosh21a,Ghosh21c,GhoshPRB2020,GhoshPRB2022,Ghosh_2024_JPCM,Ghosh2024_Flq_Anderson,GhoshAnnica_2024_}. Interestingly, an external time-periodic drive can generate topologically protected Majorana modes both at zero and finite quasi-energy, namely $0$- and $\pi$- Floquet Majorana end modes (FMEMs) respectively, even starting from nontopologcal phase of a system. Generation of FMEMs and their $dI/dV$ signature by periodically driving the static TSCs is investigated in the literature~\cite{PeterHanggiFloquet,ThakurathiPRB2013,ThakurathiPRB2017,PotterPRX2016, ThakurathiPRB2017,KunduPRL2013,Pereg_PRL_2015,Pereg_PRB_2016,MitraPRB2019,Mondal2023_NW,Mondal_2023_Shiba,Mondal2024} . However, effect of external time periodic drive in AM-SC heterostructures remains unexplored till date. Furthermore, signature of FMEMs and dynamical version of AZMs via JC response still remains an interesting avenue to address. Very recently, the JC response of FMEMs has been investigated for driven JJ of 1D $p$-wave Kitaev chain~\cite{KunduPRL2024}. 

Given the above background, in this letter, we pose the following key questions: (i) Is it possible to transform the static trivial AZMs to topological accidental zero modes
(TAZMs) with drive? (ii) In the static case, can MZMs be distinguished from AZMs via JC signatures?, and (iii) In the driven scenario, can TAZMs be identified utilizing energy resolved Floquet JC formalism? 

\begin{figure}
	\includegraphics[scale=0.35]{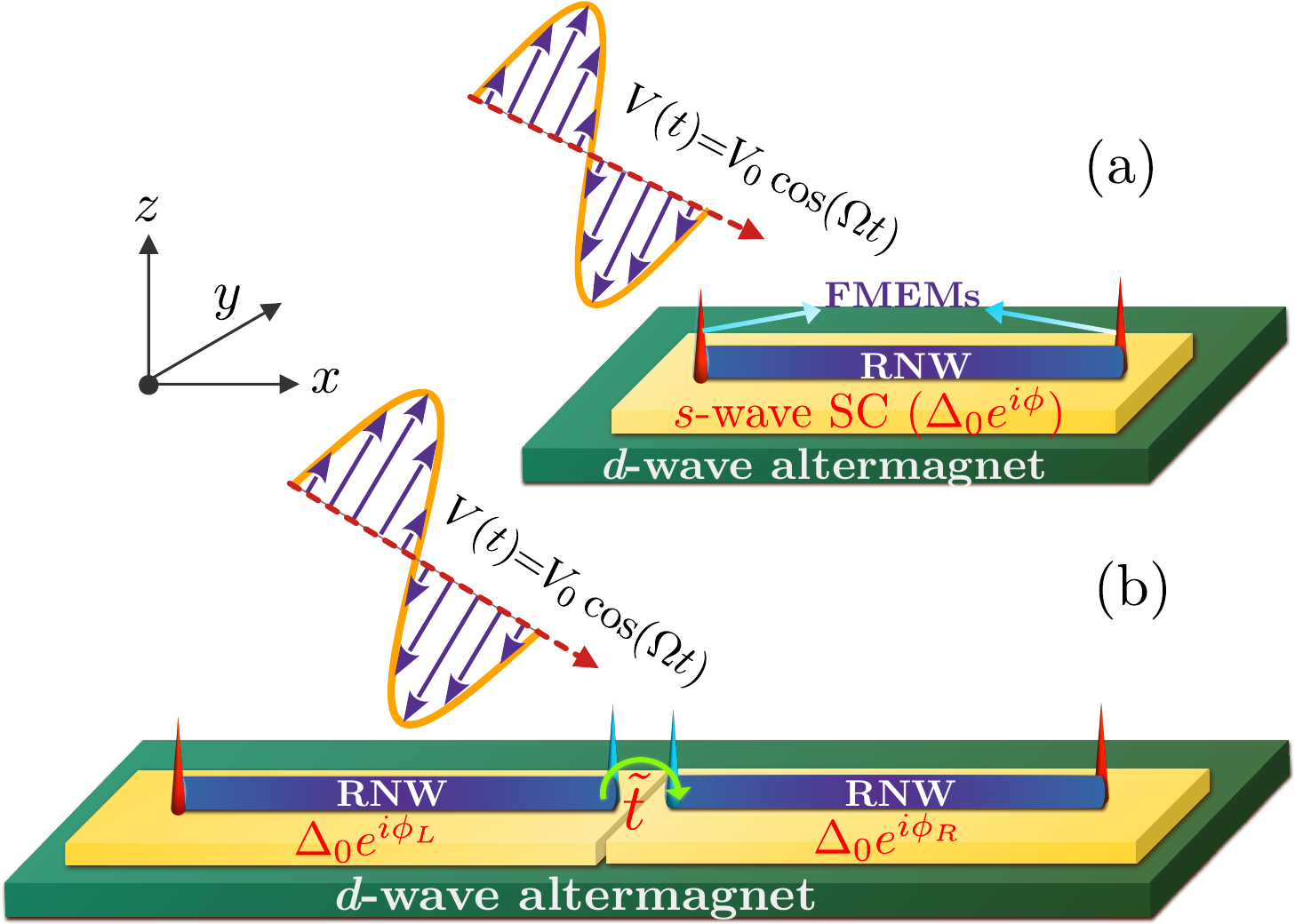}
	\caption{ \textbf{Schematic illustration of our AM-SC heterostructure:} (a) A 1D RNW (blue) is placed in close proximity to a $s$-wave SC (yellow) and a $d$-wave AM (green) in presence of an external sinusoidal Floquet drive, $V(t)$. This setup hosts FMEMs localized at the ends of the RNW (red spikes). Panel (b) depicts the JJ setup comprised of two weakly coupled TSCs (as shown in panel (a)), with coupling strength $\tilde{t}$ and superconducting phase bias, $\phi=(\phi_L-\phi_R)$. This setup is periodically driven by $V(t)$.}
	\label{Fig.1}
\end{figure}

To answer these intriguing questions, we consider a heterostructure consisting of a 1D RNW proximitized to a $d$-wave AM and $s$-wave SC, and drive the system with a sinusoidal time-periodic gate voltage, $V(t)$, as illustrated in Fig.\,\ref{Fig.1}(a). First, driving the nontopological phase of the static model, we generate both $0$- and $\pi$-FMEMs and topologically characterize the emergent Floquet TSC phase using dynamical winding numbers (DWNs) . Importantly considering the model parameters that support AZMs in the static system~\cite{Mondal2025PRBL}, we demonstrate AZMs 
can be fully gapped out by suitably tuning the driving frequency and amplitude. Interestingly, this procedure not only eliminates AZMs but also generates robust TAZM manifested as $\pi$-FMEMs. Afterwards, we construct a JJ as shown in Fig.\,\ref{Fig.1}(b) formed by two weakly coupled TSCs with a superconducting phase bias, $\phi$. 
For the undriven JJ, the junction localized MZMs display the unusual $4\pi$-periodic Josephson effect, manifested as discontinuity in JC at $\phi=\pi$ owing to its topological origin. In contrast, AZMs 
do not show such feature and exhibit $2\pi$-periodicity in JC signifying their nontopological nature. Interestingly, the TAZMs, obtained by driving the static AZMs, exhibit the $4\pi$-periodic Josephson effect indicating the topological character of them induced by the external drive. 

\textit{\textcolor{blue}{Static model Hamiltonian ---}}  We begin with the static Bogoliubov de Gennes (BdG) Hamiltonian in real space representing the 1D RNW of length, $N_x$ hosting proximity induced $s$-wave superconductivity and the magnetic order of a $d$-wave AM as shown in Fig.\,\ref{Fig.1}(a) \cite{Ghorashi2024PRL,Mondal2025PRBL},
\begin{eqnarray}
	\mc{H}_0 = \sum_{x,x'=1}^{N_x}\Psi_x^\dagger H_{x,x'} \Psi_x' + \mathrm{h.c.}\ , \non 
\end{eqnarray}
where,  %
\begin{eqnarray}
	H_{x,x'} =& \frac{1}{2}[t \, \pi_z\sigma_0   -i\lambda_R\, \pi_z \sigma_y + J_A\, \pi_0 \sigma_z] \delta_{x',x+1}\non \\ 
	& [-\mu \,\pi_z\sigma_0 +  \Delta_0 e^{i\phi} \,\pi_x \sigma_0]\delta_{x',x}\ ,
	\label{Eq.static_Ham}
\end{eqnarray}
Here, the BdG basis is chosen as: $\Psi_x = (c_{x\ua},c_{x\da},-c^\dagger_{x\da},c^\dagger_{x\ua})^T$, with 
$c_{xs} (c^\dagger_{xs})$ denotes the annihilation (creation) operator of an electron at site, $x$ with spin, $s=(\ua,\da)$. The model parameters $t,\mu,\lambda_R,J_A$ represent the strength of hopping amplitude, chemical potential, Rashba SOC, $d$-wave  altermagnetic strength,  respectively, while  $\Delta_0$ and $\phi$ correspond to the proximity induced superconducting gap and phase. The Pauli matrices, $\vect{\pi}$($\vect{\sigma}$) act on the particle-hole (spin) subspace. The static model in Eq.~\eqref{Eq.static_Ham} belongs to the BDI topological class \cite{ChiuRMP2016} as chiral ($\mc{S}$), charge-conjugation ($\mc{C}$), and pseudo- time reversal symmetry ($\mc{T}'$) are preserved~\cite{Mondal2025PRBL,TewariPRL2012} with $\mc{S}=\pi_y \sigma_y$  $\mc{C}=\pi_y \sigma_y \mc{K}$, and $\mc{T}'=\mc{K}$ where $\mc{K}$ is the complex conjugation operator. The static Hamiltonian hosts MZMs for $\dis{\sqrt{(t-\mu)^2 + \Delta_0^2} \le |J_A| \le \sqrt{(t+\mu)^2 + \Delta_0^2}}$ and AZMs for $\mu=0, |J_A|\ge t$ ~\cite{Mondal2025PRBL}. For the rest of the paper, we set $t=1$, $\Delta_0=0.3t$, and $\lambda_R=0.5t$ for simplicity while other model parameter values are explicitly mentioned. In Fig.~\ref{Fig.1}(a) we choose $\phi=0$ while allow $\phi$ to take nonzero value in the JJ setup (see Fig.~\ref{Fig.1}(b))

\textit{\textcolor{blue}{Generation and topological characterization of FMEMs---}}We periodically drive the static Hamiltonian to realize FMEMs using a time-periodic sinusoidal modulation to the chemical potential in Eq.\,\eqref{Eq.static_Ham} as, 
\begin{equation}
	\mu(t)  = \mu + V_0 \cos (\Omega t)\ , \non 
\end{equation}
Therefore, the time-periodic part of the driven system is then expressed in the BdG basis as,
\begin{equation}
	V(t) = \sum_{x=1}^{N_x} \Psi_x^\dagger [V_0 \cos (\Omega t) \pi_z \sigma_0] \Psi_x ,
	\label{Eq.Floquet Drive}
\end{equation}
where, $V_0$ and $\Omega\,(=2\pi/T)$ are the amplitude and frequency of the Floquet drive, respectively, with $T$ being the time period of the drive. The time periodicity of $V(t)$, i.e, $V(t+T) = V(t)$ renders the full Hamiltonian, $\mc{H}(t)= \mc{H}_0 + V(t)$ time-periodic with period $T$. Utilizing the Floquet theorem, the solution of the time dependent Hamiltonian can be expressed as  
$\ket{\psi_\alpha(t)} = e^{-iE_\alpha t} \ket{\varphi_\alpha(t)}$ where $\{E_{\alpha}\}$ and $\{\ket{\varphi_{\alpha}(t)}\}$ are the quasienergies and Floquet eigenstates of the system~\cite{Eckardt2017,Eckardt_2015,Oka2019}. For a time span large compared to a single time period, 
$T$, the quasienergies, $\{E_\alpha$\}, are procured in a stroboscopic fashion by diagonalizing the effective time-independent Floquet Hamiltonian, $\mc{H}_F$, defined as~\cite{Eckardt_2015},
\begin{equation}
	\mathrm{exp}(-i \mc{H}_F T) \equiv  \!\mathrm{\hat{TO} \, exp}[-i\!\! \int_{0}^{T}\!\! \mc{H}(t) dt]\ ,
\end{equation}
where, $\mathrm{\hat{TO}}$ corresponds to time-ordering operator. We numerically compute the quasienergies, $\{E_\alpha\}$ efficiently employing the second-order Trotter-Suzuki formalism~\cite{Suzuki1976,Raedt1983_Suzuki_Trotter,Tao2022_Suzuki_Trotter,Ghosh_2024_JPCM}. 
Note that, in real-time picture $E_\alpha$\,s are restricted to be in the range  $[-\Omega/2,\Omega/2]$, known as first Floquet Brillouin zone. Following the literature, we call the Floquet modes with energy $E_\alpha=0$ and $\Omega/2$ as $0$-FMEMs and $\pi$-FMEMs, respectively~\cite{Mondal2023_NW,Mondal_2023_Shiba,Ghosh_2024_JPCM}.
%
\begin{figure}
	\includegraphics[scale=0.39]{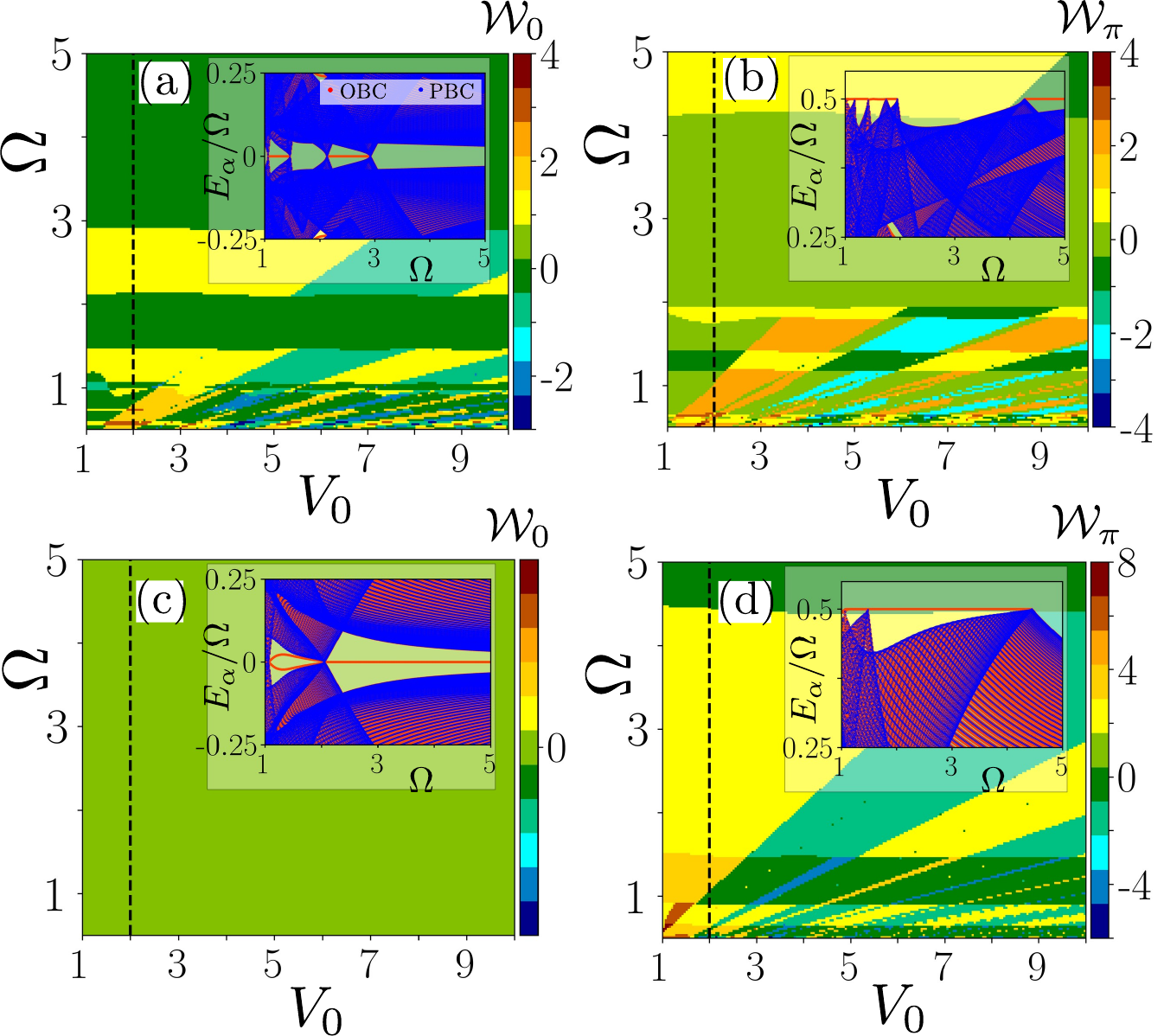}
	\caption{ \textbf{Illustration of DWNs and quasienergy spectrum:} In panels (a)-(b), we depict the DWNs, $\Wz$ and $\Wp$ in the $V_0\mhyphen\Omega$ plane when the static model is in the trivial phase and does not host any zero energy modes, while panels (c)-(d) showcase the same when the static model hosts AZMs (see SM \cite{supp} for static winding number of the undriven system). Inset of each panel displays the quasienergy spectrum, $E_\alpha/\Omega$ as a function of $\Omega$ (along the vertical dashed lines of each panel) for $V_0=2$ employing both OBC and PBC, highlighting the emergence of $0$-FMEMs (inset of (a),(c)) and $\pi$-FMEMs (inset of (b),(d)). We choose the model parameters as $(\mu/t,J_A/t=1.5,0.4)$ in panels (a)-(b) and $(\mu/t,J_A/t=0,1.2)$ in panels (c)-(d) while consider a finite size system with $N_x=100$ lattice sites across all the panels. }
	\label{Fig.2}
\end{figure}
For the purpose of topological characterization of 0- and $\pi$-FMEMs, we compute the DWNs, $\Wz$ and $\Wp$ exploiting the chiral symmetry, $\mc{S}$, of the system (See SM ~\cite{supp} for detailed derivation) as~\cite{Asboth2014,Benalcazar2022_PRL,Ghosh2024_Flq_Anderson},
\begin{equation}
	\Wz = \frac{\nu_1 + \nu_2}{2} \,\, \mathrm {\,and\,}\,\, \Wp = \frac{\nu_1 - \nu_2}{2}\ ,
\end{equation} where, 
\begin{equation*}
	\nu_m = \frac{1}{2\pi i} \mathrm{Tr\, ln} (\bar{P}^m_A\bar{P}_B^{m\dagger}) \in \mathbb{Z} \mathrm{,\,\, with\,\,} m=1,2
\end{equation*}

Here, $\bar{P}^m_{A,B}$ represents the polarization operator restricted to the sublattices $A,B$ and defined as $\bar{P}^m_{A,B} =U^{m\dagger}_{A,B} U^\dagger_{A,B} P_x U_{A,B}U^m_{A,B}$  with $P_x= \sum_{x=1,\alpha}^N c^\dagger_{x,\alpha} \mathrm{exp}(-2\pi i\, x/N_x) c_{x,\alpha}$ \cite{Resta1998}. Note that, $U^m_{A,B}$ and $U_{A,B}$ denote the unitary matrices constructed from the real space eigenvectors of $\mathcal{H}_F$ and $\mc{S}$ respectively. 

We implement the periodic drive into two topologically trivial regions of the static phase diagram (see SM\,\cite{supp} for the static phase diagram and number of zero energy end localized modes): (i) one without the presence of any zero-energy modes, and (ii) one hosting only nontopological AZMs~\cite{Mondal2025PRBL}. In Figs.~\ref{Fig.2}(a) and \ref{Fig.2}(b), we illustrate the variation of $\Wz$ and $\Wp$ in the $V_0\mhyphen\Omega$ plane for the case (i), where nonzero integer values of $\Wz$ and $\Wp$ indicate the number of $0$- and $\pi$-FMEM pairs, respectively. In the low frequency regime, the driven system hosts multiple FMEMs ($\Wz, \Wp > 1$), as the static model band width is much larger than the width of the first Floquet zone. To corroborate with the DWNs, we compute the quasienergy spectrum $\{E_\alpha\}$ by diagonalizing the Floquet Hamiltonian, $\mc{H}_F$, under both open (OBC) and periodic (PBC) boundary conditions. The insets of Figs.~\ref{Fig.2}(a) and \ref{Fig.2}(b) display $E_\alpha/\Omega$ as a function of $\Omega$ for $V_0 = 2$, confirming the emergence of $0$- and $\pi$-FMEMs at $E_\alpha/\Omega = 0$ and $0.5$, respectively. Specifically, $\Wz=1$ and $\Wz=2$ correspond to two and four 0-FMEMs, respectively, localized at the ends of the NW as confirmed by the eigenvalue spectrum and site-resolved local density of states (see SM\,\cite{supp} for details). These findings establish the generation of $0$- and $\pi$-FMEMs driving the trivial static phase of the AM-SC heterostructure.

To investigate the aftermath of nontopological AZMs~\cite{Mondal2025PRBL} under periodic drive, we depict the variation of $\Wz$ and $\Wp$ in Figs.\,\ref{Fig.2}(c) and (d), respectively starting from the regime (ii) of the static model. In Fig.\,\ref{Fig.2}(c), $\Wz$ is zero for any values of $V_0$ and $\Omega$ implying the absence of topological $0$-FMEMs. Notably, for $V_0=2$ and $\dis{1<\Omega<2}$ of the drive, zero energy AZMs are gapped out as shown in the inset of Fig.\,\ref{Fig.2}(c). However, AZMs still persist for large $\Omega$, being the reminiscent of AZMs in the static Hamiltonian. Despite the absence of $0$-FMEMs, in Fig.\,\ref{Fig.2}(d) we observe $\Wp$ to inherit even integer values implying the presence of multiple $\pi$-FMEMs, which we coin as `TAZMs' as they emerge by driving the trivial AZMs. The quasienergy spectrum, $E_\alpha/\Omega$, depicted as a function of $\Omega$, clearly reveals the appearance of $\pi$-FMEMs at $E_\alpha/\Omega=0.5$ (see inset of Fig.\ref{Fig.2}(d)). These results confirm that Floquet driving can eliminate nontopological AZMs while inducing robust topological $\pi$-FMEMs, addressing the first key question of our study. 

For a comprehensive analysis, we calculate the bulk gap associated with $0$- and $\pi$-FMEMs, which supports DWNs phase diagram (see SM~\cite{supp} for details). Moreover, we investigate the robustness of DWNs against static onsite random disorder and find that $\pi$-FMEMs turn out to be more robust compared to the $0$-FMEMs (see SM~\cite{supp} for further details).


\textit{\textcolor{blue}{Josephson current signature of static MZMs/AZMs -}} To answer the second question, we consider a JJ formed by two weakly coupled TSCs with coupling strength, $\tilde{t}$, and a phase difference, $\dis{\phi_L \!-\!\phi_R=\phi}$, between them as shown in Fig.\,\ref{Fig.1}(b). Here, each TSC corresponds to our 1D RNW with proximity induced AM and SC. First, we switch off the periodic drive and investigate the JC in the static case to distinguish between MZMs and AZMs 
as proposed in the static Hamiltonian~\cite{Mondal2025PRBL}. The Hamiltonian of the JJ is constructed as, 
\begin{equation}
	\dis{\mc{H}_{\jj} =\mc{H}_0^L + \mc{H}_0^R + \mc{H}^{LR}}\ ,
\end{equation}
where,\begin{eqnarray}
	\mc{H}_0^{\alpha} &=&  \sum_{ x,x'=1}^{N_x}  \Psi_x^{\alpha\dagger} H_{x,x'}^\alpha(\phi_\alpha)  \Psi_{x'}^{\alpha} ,  \mathrm{\,\, \,\,\,}\alpha=L,R \non \mathrm{\,\,and,\,\,} \\
	\mc{H}^{LR} &=& \Psi_{N_x}^{L\dagger} \left[\frac{\tilde{t}}{2}\,\, \pi_z\sigma_0 \right] \Psi_{1}^{R} + \mathrm{h.c}\ .
\end{eqnarray}
Here, $\Psi_x^\alpha$($\Psi_x^{\alpha\dagger}$) corresponds to the annihilation (creation) operator of the electron at site $x$ in the left ($\alpha=L$) and right ($\alpha=R$) SC. Also, $H_{x,x'}^{L/R}(\phi_{L/R})$ represents the Hamiltonian for the left/right TSCs, governed by the Eq.\,\eqref{Eq.static_Ham} with superconducting phase $\phi_{L/R}$. The third term, $\mc{H}^{LR}$, denotes the coupling between the last site of left SC and first site of the right SC. For the purpose of our study, we consider two situations when each TSC host either topological MZMs or nontopological AZMs at its ends. 

We compute the JC, $I(\phi)$, flowing across the junction in presence of phase bias, $\phi$, following the relation~\cite{Beenakker1991,Krishendu2004_Josephson},
\begin{equation}
	I(\phi) = \sum_{\beta} f_{\rm FD}(\epsilon_\beta)\frac{\partial \eps_\beta(\phi)}{\partial \phi}\ , \label{Eq.JosephsonCurrent}
\end{equation}
where, $f_{\rm FD}(\eps_\beta)$ denotes the Fermi-Dirac distribution function representing the occupation probability of the $\beta^{\rm th}$ energy eigenstate with energy $\eps_\beta$. In the zero temperature limit, the summation in Eq.\,\eqref{Eq.JosephsonCurrent}, involves only the negative energy eigenvalues \ie $\eps_\beta\le 0$.  

\begin{figure}
	\includegraphics[scale=0.33]{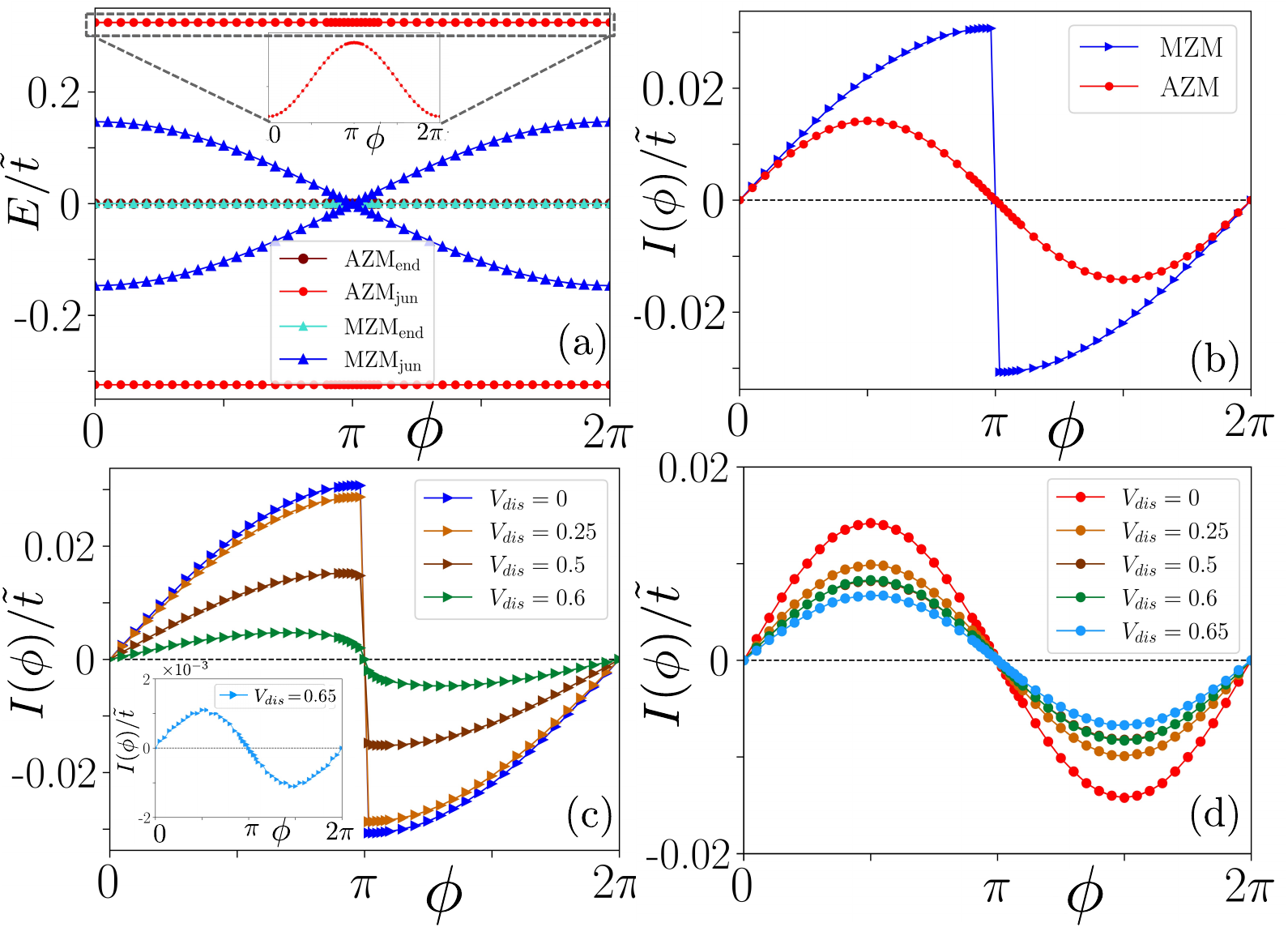}
	\caption{\textbf{Behaviour of JC flowing across the static JJ:} In panel (a), we present the $\eps-\phi$ relation due to MZMs and AZMs localized near (${\rm AZM_{jun}, MZM_{jun}}$) and far from the junction (${\rm AZM_{end}, MZM_{end}}$).
		In panel (b), we depict the variation of JC, $I(\phi)$ as a function of $\phi$ for MZMs and AZMs. Panels (c) and (d) display the effect of disorder on $I(\phi)$ for different disorder strengths, $V_{dis}$ corresponding to MZMs and AZMs, respectively.  For MZMs, we choose $(\mu, J_A) = (1.1t, 0.4t)$, while for AZMs $(\mu, J_A) = (0, 1.2t)$. We fix the system size $N_x = 340$ lattice sites and $\tilde{t} = 0.01t$ in all panels. We consider 50 disorder configurations in panels (c) and (d).}
	\label{Fig.3}
\end{figure}

We investigate the influence of phase bias $\phi$ on $ \eps_\beta(\phi)$ for MZMs and AZMs localized near and far from the junction in Fig.~\ref{Fig.3}(a). Due to the coupling, $\tilde{t}$, between the TSCs, the modes localized near the junction (blue spikes in Fig.\,\ref{Fig.1}(b)), hybridize forming finite energy quasiparticle states. In contrast, the zero modes far from the junction (red spikes in Fig.\,\ref{Fig.1}(b)), remains at zero energy as shown in Fig.~\ref{Fig.3}(a), provided $\dis{N_x\gg\xi_M}$ where $\xi_M$ being the Majorana localization length. Interestingly, the junction localized MZMs follow $\eps^{\rm jun }_\pm (\phi)\propto \pm \cos (\phi/2)$ leading to energy crossing at $\phi=\pi$ which signals a switching of the ground-state fermion parity resulting in the emergence of  $4\pi$-periodic Josephson effect~\cite{Alicea_2012,Alicea2011_NatPhys,Beenakker2013search}. In sharp contrast, junction localized AZMs, though $\phi$ dependent (see inset of Fig.~\ref{Fig.3}(a)), do not exhibit any energy crossing which indicates no change in fermion parity and confirming their nontopological nature, although predicted to be topological in Ref.~\cite{Ghorashi2024PRL}. In Fig.\,\ref{Fig.3}(b), 
we depict the behavior of static JC $I(\phi)$ for both AZMs and MZMs where for MZMs, a clear discontinuity at $\phi = \pi$ reflects $4\pi$ periodicity, while for AZMs, $I(\phi)$ is $2\pi$-periodic with a maximum at $\phi = \pi/2$. This distinct behaviour in JC offers a clear signature to distinguish the AZMs from MZMs (apart from $dI/dV$ and shot noise signature~\cite{Mondal2025PRBL}), thus addressing our second key question. We also introduce a random static onsite disorder in the Hamiltonian, $\mc{H}_{dis} = \sum_{x} \Psi_x^\dagger [V(x) \pi_z\sigma_0] \Psi_x$ where $V(x)$ is a random number uniformly distributed in the range $[-V_{\rm dis}/2,V_{\rm dis}/2]$. We present the effect of disorder on the JC for both MZMs and AZMs in Fig.\,\ref{Fig.3}(c) and Fig.\,\ref{Fig.3}(d), respectively. For MZMs, the discontinuity at $\phi=\pi$ diminishes with increasing disorder strength and for strong disorder, the $4\pi$ periodic nature is governed by conventional $2\pi$-periodic JC (see inset of Fig.\,\ref{Fig.3}(c)). On the other hand, in case of AZMs, disorder suppresses the JC amplitude without altering its periodicity.

\textit{\textcolor{blue}{Josephson current signature of FMEMs/TAZMs ---}} Here, we apply identical time-periodic drive $V(t)$ (Eq.~\eqref{Eq.Floquet Drive}) to both the TSCs in the JJ setup as illustrated in Fig.\,\ref{Fig.1}(b) to investigate Floquet JC signatures of FMEMs. The Floquet drive carries the system into a out-of-equilibrium phase where the occupation of quasienergy states deviates from the equilibrium distribution. When such a driven system is weakly coupled to a thermal reservoir at temperature $\theta_r$ and chemical potential $\mu_r$, the steady-state occupation of a quasienergy state, $E_\alpha$, is governed by \cite{KunduPRL2024}, 
\begin{equation}
	n( E_\alpha,\mu_r) = \sum_{m\in\mathbb{Z}} \braket{u_\A^{(m)}|u_\A^{(m)}} f_{\rm FD}( E_\A + m\Omega - \mu_r)\ . 
	\label{Eq.Occupation prob}
\end{equation}
where, $\ket{u_\A^{(m)}}$'s are the Fourier modes of $\ket{\varphi_{\A}(t)}$ \ie $\ket{\varphi_{\A}(t)}=\sum_{m\in\mathbb{Z}} e^{-im\Omega t}\ket{u_\A^{(m)}}$. Importantly, Eq.~\eqref{Eq.Occupation prob} allows energy resolved probing of the $0$- and $\pi$-FMEMs by setting $\mu^0_r = n\Omega$ and $\mu^\pi_r = (n+1/2)\Omega$, respectively with $n\in \mathbb{Z}$. The JC, averaged over the full time cycle, is computed utilizing the relation \cite{KunduPRL2024} (see SM~\cite{supp} for brief derivation),
\begin{equation}
	I_{0/\pi}(\phi) = \sum_\A n(E_\A,\mu^{0/\pi}_r) \frac{\partial E_\A(\phi)}{\partial \phi}\ , \label{Eq.FloquetJC}
\end{equation}
Note that, performing brute force derivatives of $E_\A(\phi)$, without invoking the correct occupation number, may produce a discontinuous jump in JC at $\phi=\pi$ \cite{Roy2025_FlqJJ}, but fails to distinguish between the presence of $0$- and $\pi$-FMEMs (See SM \cite{supp} for further details). Crucially, even when FMEMs are present, the $4\pi$-periodicity in $I(\phi)$ emerges only if $N_x \gg \xi_m$. Otherwise, a conventional $2\pi$-periodic Josephson effect is restored due to overlap between the modes localized near and far from the junction (see SM~\cite{supp} for details). 
%
\begin{figure}
	\includegraphics[scale=0.35]{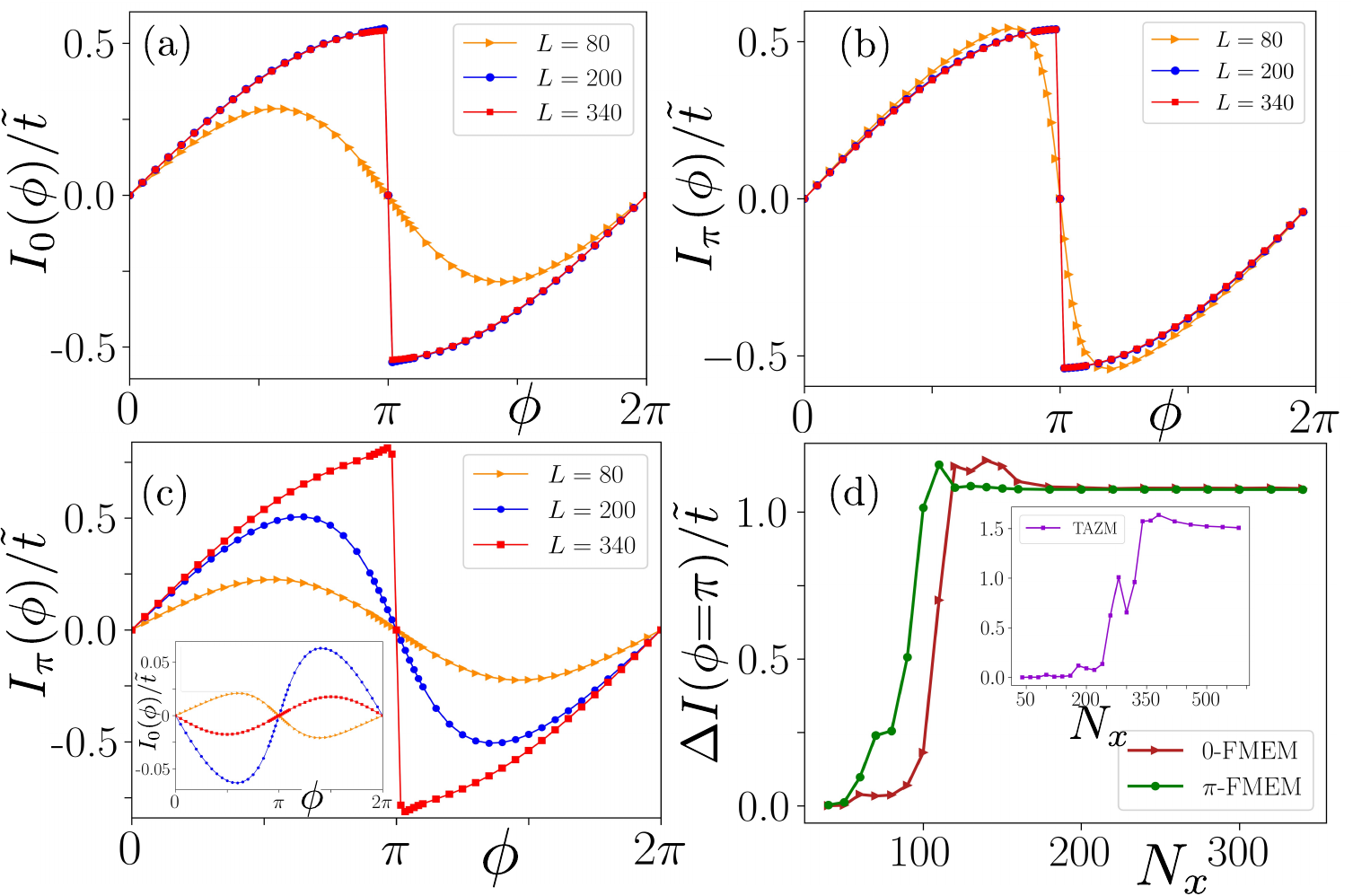}
	\caption{\textbf{Variation of Floquet JC across the driven JJ:} In panels (a) and (b), we present the variation of JC when the driven system hosts only $0$-FMEMs and only $\pi$-FMEMs, respectively. On the other hand, panel (c) [inset] illustrates the JC for TAZMs [zero energy Floquet AZMs] for various system sizes. Panel (d) highlights the discontinuous jump in JC at $\phi=\pi$ as a function of $N_x$ corresponding to panels (a) and (b) while the inset represents the discontinuity for TAZMs as presented in panel (c). The model parameters in $(\mu,J_A,\Omega,V_0)$ space is chosen as, (a) $(1.5t,0.4t,2.4,4.2t)$,(b) $(1.5t,0.4t,4.8,4.5t)$, (c) $(0,1.2t,3,3.1t)$ and $\tilde{t}=0.01t$ in all the panels.}
	\label{Fig.4}
\end{figure}

We compute $I_{0/\pi}(\phi)$ corresponding to three phases of the driven system hosting (i) only $0$-FMEMs (Fig.~\ref{Fig.4}(a)), (ii) only $\pi$-FMEMs (Fig.~\ref{Fig.4}(b)), and (iii) multiple $\pi$-FMEMs (TAZMs) with nontopological AZMs at $E_\A=0$ (Fig.~\ref{Fig.4}(c)) considering various lengths of TSCs. In cases (i) and (ii), we observe the $4\pi$-periodic Josephson effect in $I_0 (\phi)$ and $I_\pi(\phi)$ respectively when $N_x\gg \xi_M$. Interestingly, in case (iii) only $I_\pi(\phi)$  exhibits a discontinuous jump at $\phi=\pi$, however $I_0(\phi)$ remains $2\pi$-periodic (see inset of Fig.~\ref{Fig.4}(c)) even if we increase the length of the two TSCs, highlighting the nontopological nature of Floquet AZMs that remain at high frequency. The crossover from $2\pi$ to $4\pi$ periodicity is clearly visible with increasing the system size as captured by the discontinuos jump in $\Delta I(\phi=\pi)$ (see Fig.\,\ref{Fig.4}(d)). The critical length $N^c_x$ for such crossover, is governed by the Majorana localization length $\xi_M$. Although, the later varies with model parameters, for cases (i) and (ii), $N^c_x \simeq 100$ while for TAZMs, $N^c_x \simeq 300$. Physically, such different length scales can be attributed to the corresponding size of $\pi$-gap protecting the FMEMs and TAZMs. Additionally, the $4\pi$-periodic Josephson effect can only distinguish the topological modes from the nontopological ones. However, it cannot differentiate between two topological phases \eg between $\Wz=1$ and $\Wz=2$ with any universal feature, since in both cases JC remains $4\pi$-periodic (see SM\,\cite{supp} for details) via the discontinous jump at $\phi=\pi$.  These results establish that the TAZMs, obtained by periodically driving the phase hosting AZMs exhibit the hallmark $4\pi$-periodic Josephson effect, answering our final key question.

\textit{\textcolor{blue}{Summary and Discussion.---}} 
To summarize, in this article, we periodically drive a heterostructure consists of a 1D RNW proximitized to a bulk $s$-wave SC and $d$-wave AM to generate FMEMs from the nontopological regime of the static phase. Topological characterization employing DWNs $\mc{W}_0$ and $\mc{W}_\pi$ reveals a rich phase diagram hosting multiple $0$- and $\pi$-FMEMs. Crucially, nontopological AZMs can be gapped out by the drive, while inducing multiple $\pi$-FMEMs, termed as TAZMs, inheriting topological properties. Note that, in the high frequency limit, the system hosts nontopological zero energy AZMs being the reminiscent of the static AZMs.  To distinguish between MZMs and AZMs, we choose a JJ setup of two weakly coupled TSCs. While MZMs exhibit a $4\pi$-periodicity in JC due to fermion parity switching at $\phi = \pi$, for AZMs JC remains $2\pi$-periodic, confirming their nontopological nature. Extending this to the driven system, we use energy-resolved Floquet JC formalism~\cite{KunduPRL2024} to identify $0$- and $\pi$-FMEMs via $4\pi$-periodic signatures. Intriguingly, TAZMs also display discontinuities in the JC, revealing $4\pi$ periodicity at sufficiently larger system sizes owing to large localization length (smaller $\pi$-gap). Our results establish a robust framework to engineer, detect, and distinguish dynamical topological modes utilizing Floquet JC. Generation of FMEMs in planar Josephson junctions using a time-periodic gate voltage at the barrier region was reported in Ref.\,\cite{MitraPRB2019}. 
However, instead of computing JC, they relied on $\rm{dI/dV}$ and density of states to probe the FMEMs.

In addition to better proximity effect due to zero net magnetization, the use of AM instead of external Zeeman field can offer further advantages in both static and driven systems. In static case, the bulk topological superconducting gap is significantly enhanced in altermagnetic heterostructure, sometimes even twice of the bulk gap obtained with Zeeman field, providing better stability against disorder in comparision with the heterostructure with external magnetic field. Moreover, enhnaced bulk gap leads to the enhanced temperature range, below which Majorana modes are possible to observe in experiments. Furthermore, in driven systems, AM extends the topological phase boundary beyond the phase boundary obtained with Zeeman field leading to enlarged parameter space for the realization of FMEMs. We refer to the  SM\,\cite{supp} for the detailed discussion.

As far as experimental detection of $4\pi$-periodic Josephson effect is concerned, this may face key challenges: (i) quasiparticle poisoning, where external tunneling processes from attached leads can alter the fermion parity and restore the conventional $2\pi$-periodicity~\cite{Beenakker2013search,Rokhinson2012}, and (ii) Landau-Zener tunneling wherein a highly transparent $2\pi$-periodic state can tunnel through the avoided crossing at $\phi=\pi$, mimicking the $4\pi$-periodic Josephson effect~\cite{Billangeon2007_LZtunneling,Wiedenmann2016}. To circumvent these challenges, experimental detection often relies on the a.c. Josephson effect, where the appearance of only even-integer Shapiro steps indicate topological superconductivity~\cite{Rokhinson2012,Wiedenmann2016}. However, experimental progress to detect FMEMs in periodically driven JJ still remains in its infancy~\cite{Park2022}. Concerning the realistic physical microscopic model parameters, in our calculations we have scaled all the energy scales in units of strength of hopping amplitude, $t$, instead of considering actual values corresponding to real materials. 
However, we propose an estimate of the parameter values from our numerical simulation (not based on real materials) for which the intriguing features of our system appear. A representative set of parameter values are: $t \simeq 1 \,{\rm meV}$ (suppose given for any NW material), then other parameters are $\; \Delta_0 \simeq 300\,{\rm \mu eV}, \; \lambda_R \simeq 500\,{\rm \mu eV}, \; J_A = 0.5\text{--}1\,{\rm meV}, \; V_0 \leq 2.5\,{\rm meV},$ and $\Omega \sim 40\,{\rm GHz}$.
Moreover in future, one can explore a.c. Josephson signatures in driven systems to distinguish trivial zero modes from topological Majorana modes. Furthermore, employing numerous driving protocols, one may also explore the possibility of complete elimination 
of AZMs from this hybrid setup.

\textit{\textcolor{blue}{Acknowledgments---}} We acknowledge Arijit Kundu and Rekha Kumari for stimulating discussions. A.P., D.M. and A.S. acknowledge SAMKHYA: High-Performance Computing Facility provided by Institute of Physics, Bhubaneswar and the two workstations provided by Institute of Physics, Bhubaneswar from DAE APEX Project, for numerical computations. T.N. acknowledges NFSG from Grant No. BITS Pilani NFSG/HYD/2023/H0911. 

\textit{\textcolor{blue}{Data Availability Statement---}} The datasets generated and analyzed during the current study are available from the corresponding author upon reasonable request.

\bibliographystyle{apsrev4-2mod}
\bibliography{bibfile.bib}

\newpage


\normalsize\clearpage
\begin{onecolumngrid}
	\begin{center}
		{\fontsize{12}{12}\selectfont
			\textbf{Supplementary Material for ``Josephson current signature of Floquet Majorana  and topological accidental zero modes in altermagnet heterostructures''\\[5mm]}}
		{\normalsize  Amartya Pal\orcidA,$^{1,2}$ Debashish Mondal\orcidD{},$^{1,2}$ Tanay Nag\orcidB{},$^{3}$ and  Arijit Saha\orcidC{},$^{1,2}$ \\[1mm]}
		{\small $^1$\textit{Institute of Physics, Sachivalaya Marg, Bhubaneswar-751005, India}\\[0.5mm]}
		{\small $^2$\textit{Homi Bhabha National Institute, Training School Complex, Anushakti Nagar, Mumbai 400094, India}\\[0.5mm]}
		{\small $^3$\textit{Department of Physics, BITS Pilani-Hydrabad Campus, Telangana 500078, India}\\[0.5mm]}
		{}
	\end{center}
	
	\newcounter{defcounter}
	\setcounter{defcounter}{0}
	\setcounter{equation}{0}
	\renewcommand{\theequation}{S\arabic{equation}}
	\setcounter{page}{1}
	\pagenumbering{roman}
	
	\renewcommand{\thesection}{S\arabic{section}}
	
	\tableofcontents 
	
\section{Real space formalism of dynamical winding number (DWN)} 
\label{sec:DWN_Formalism}

\begin{figure}[h]
	\centering
	\includegraphics[scale=0.6]{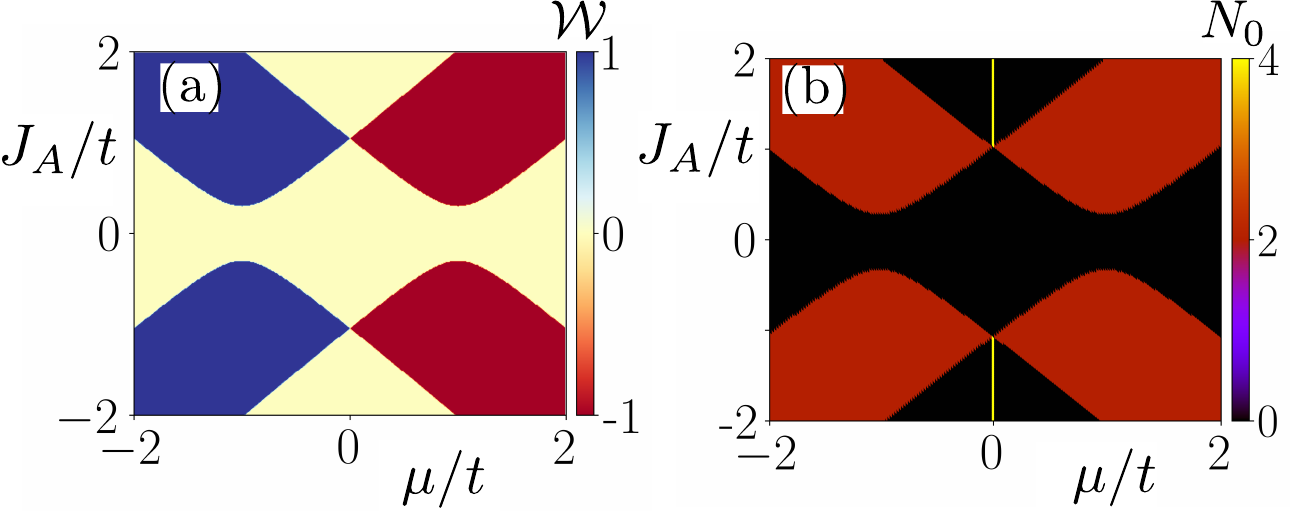}
	\caption{\tbf{Variation of winding number $\mc{W}$ and number of zero modes $N_{0}$ for the static case:} In panel (a) we depict the static winding number $\mc{W}$ in the $\mu/t \mhyphen J_A/t$ plane for the undriven system (see Eq.(1) of the main text). In panel (b), we highlight the number of zero energy modes, $N_0$, localized at the ends of the AM based 1D TSC. Note that in panel (b) $N_0=4$ for 
		$\mu/t=0$ and $|J_A/t|\ge 1$ signifying the presence of four end localized zero energy modes which we refer as AZMs. However, for the same parameter values $\mc{W}$ is zero in panel (a) which establishes their nontopological nature. Other model parameters of the static Hamiltonian are chosen as 
		$\lambda_R=0.5t, \Delta_0=0.3t$ in both the panels. }
	\label{Fig.S0}
\end{figure}
In the main text, we have utilized the DWNs, $\mc{W}_0$ and $\mc{W}_{\pi}$, to topologically characterize the $0$- and $\pi$- Floquet Majorana end modes (FMEMs), respectively generated via applying time periodic drive in the altermagnet based TSC. Here, we briefly discuss the computation of the DWNs based on a real space formalism exploiting the chiral symmetry of the system, following Refs.~\cite{Asboth2014,Benalcazar2022_PRL,Ghosh2024_Flq_Anderson}. Presence of chiral symmetry, $\mc{S}$, in the system incorporates the following constraint to the time periodic Hamiltonian: $\mc{H}(t)= \mc{H}_0 + V(t)$ (see Eq.(1) and Eq.(2) in the main text), and to the time evolution operator:  $\dis{\mc{U}(t,0)=\hat{{\rm TO}}\,\, \mc{U}(t,0)=\!\mathrm{\hat{TO} \, exp}\left[-i\!\! \int_{0}^{t}\! \mc{H}(t) dt\right]}$ as~\cite{Asboth2014,Ghosh2024_Flq_Anderson} ($\hat{{\rm TO}}$ is the time-ordering operator),   
\begin{eqnarray}
	\mc{S}^{-1}\mc{H}(t)\mc{S} = - \mc{H}(T-t), \\
	\mc{S}^{-1}\mc{U}(t,0)\mc{S} = \mc{U}(T-t,0)\,\mc{U}^\dagger(T,0)\ .
\end{eqnarray}

This allows one to define two unitary operators, $U_a = \mc{U}(T/2,0)$ and $U_b = \ch^{-1} U_a \ch = \mc{U}(T,T/2)$, dividing the full time period, $T$, into two equal parts. Using these two unitary matrices $U_{a,b}$, one can define two Floquet operators as, $U_1=U_b U_a=\mc{U}(T,0)$ and $U_2 = U_a U_b=\mc{U}(3T/2,T/2)$, representing two time evolution operator over full time period but starting from $t=0$ and $T/2$, respectively. Furthermore, from $U_1$ and $U_2$, we can define two Floquet Hamiltonians as, $\mathrm{exp}(-i \,\mc{H}_F^{1,2}\, T) = U_{1,2} $ which obey the relation $\ch^{-1}\mc{H}_F^{1,2}\ch = -\mc{H}_F^{1,2}$. For a one dimensional (1D) system with $N$ lattice sites, $\mc{H}_F^{1,2}$ becomes a $4N\times4N$ dimensional matrix (including both spin and particle hole degrees of freedom). Employing the chiral symmetry, we recast $\mc{H}_F^{1,2}$ into anti-diagonal matrices by
performing a basis transformation as, 
\begin{equation}
	U_S^\dagger\mc{H}_F^{m}U_S = \begin{pmatrix}
		0 & h^m\\
		h^{m\dagger} & 0  
	\end{pmatrix} \mathrm{,\,\,where \,\,} m=1,2.
\end{equation}
Here, $U_S$ is an unitary matrix with dimension $4N\times4N$, constructed using the eigenvectors of the chiral symmetry operator, $\ch$ i.e. $U_S \ch U_S^\dagger = {\rm diag (1,1,...1,-1,-1,...,-1)}$. The off-diagonal blocks, $h^m$ and $h^{m\dagger}$, represent $2N\times 2N$ dimensional matrices. The chiral symmetry divides the whole system into two sublattices, $A$ and $B$, with eigenvalues, $+1$ and $-1$, respectively. The spectral decomposition of $\ch$ can be written as $\dis{S= U^S_A - U^S_B }$ where, $U^S_A=\sum \ket{A}\bra{A}$ and $U^S_B=\sum \ket{B}\bra{B}$.  Now performing a singular value decomposition of $h^m$ as $h^m = \bar{U}_A^m \Sigma^m \bar{U}_B^m$, one obtains the singular vectors $\bar{U}_A^m$ and $\bar{U}_B^m$, and singular values $\Sigma^m$ with $m=1,2$. We utilize $\bar{U}_{A,B}^m $ and $U^S_{A,B}$ to define two integers as~\cite{Asboth2014,Ghosh2024_Flq_Anderson}, 
\begin{equation}
	\nu_m = \frac{1}{2\pi i} \mathrm{Tr\, ln} (\bar{P}^m_A\bar{P}_B^{m\dagger}) \in \mathbb{Z} \ .
\end{equation}

Here, $\bar{P}^m_{A}$ and $\bar{P}^m_{B}$ denote the polarization operators restricted to the sublattice $A$ and $B$, respectively, and defined as $\bar{P}^m_{A,B} =\bar{U}^{m\dagger}_{A,B} U^\dagger_{A,B} P_x U_{A,B}\bar{U}^m_{A,B} $. The polarization operator $P_x= \sum_{x=1,\alpha}^N c^\dagger_{x,\alpha} \mathrm{exp}(-2\pi i\, x/N) c_{x,\alpha}$ ($N$ = length of the 1D system) is a $4N\times 4N$ dimensional matrix. To this end, we obtain the DWNs, $\Wz$ and $\Wp$ as~\cite{Asboth2014,Ghosh2024_Flq_Anderson}, 
\begin{equation}
	\Wz = \frac{\nu_1 + \nu_2}{2} \,\, \mathrm {\,and\,}\,\, \Wp = \frac{\nu_1 - \nu_2}{2}\ .
\end{equation}
Note that, the above formalism does not demand translational symmetry of the system which allows one to study the stability of $\Wz$ and $\Wp$ in the presence of disorder (See latter text for discussion).

For completeness and better understanding of the reader, in Fig.\,\ref{Fig.S0}(a) we present the variation of the static winding number $(\mc{W})$~\cite{Mondal2025PRBL}, which topologically characterizes the 1D TSC (see Eq.~(1) of the main text) in absence of any periodic drive, as a function of altermagnetic strength $J_A$, and chemical potential $\mu$ of the undriven system. Furthermore, we highlight the emergence of static nontopological accidental zero modes (AZMs) in Fig.~\,\ref{Fig.S0}(b) at $\mu/t=0$ and $|J_A/t|\ge 1$ with $t$ being the nearest neighbour hopping amplitude~\cite{Mondal2025PRBL}. These AZMs appear in regions where $\mc{W}$ vanishes (see Fig.\,\ref{Fig.S0}(a) at $\mu/t=0$ and $|J_A/t|\ge 1$). Crucially, these modes do not arise from bulk band gap closing, justfying their nontopological origin.

\section{Quasi-energy spectra and LDOS of 0- and $\pi$-FMEMs} \label{sec:Eigenvalue_spectra_LDOS}
In Fig.\,\ref{Fig.2} of the main text, we have shown the real space quasienergy spectrum, employing both open and periodic boundary condition, as a function of frequency of the external Floquet drive, $\Omega$. Here, we explicitly highlight the differences between the number of FMEMs within the topological phases along with their localization profile in real space. In Fig.\,\ref{Fig.S1}, we depict the real-space energy eigenvalue spectrum, $E_n$, as a function of the state index $n$ (panels (a)--(d)), together with the energy-resolved site-dependent local density of states (LDOS) (panels (e)--(h)), which clearly demonstrate that the FMEMs are localized at the ends of the one-dimensional system. Panels (a) and (b) in Fig.\,\ref{Fig.S1} depict the real-space energy spectra for $\mc{W}_0=1$ (hosting two 0-FMEMs) and $\mc{W}_0=2$ (hosting four 0-FMEMs), respectively, whereas panels (c) and (d) illustrate the cases $\mc{W}_\pi=1$ (two $\pi$-FMEMs) and $\mc{W}_\pi=2$ (four $\pi$-FMEMs), respectively. Panels (e) and (f) represent the LDOS at $E=0$, corresponding to the panels (a) and (b), whereas panels (g) and (h) depict the LDOS at $E=\pm \Omega/2$ corresponding to the panels (c) and (d), respectively.
\vspace{0.2cm}
\begin{figure}
	\centering
	\includegraphics[scale=0.68]{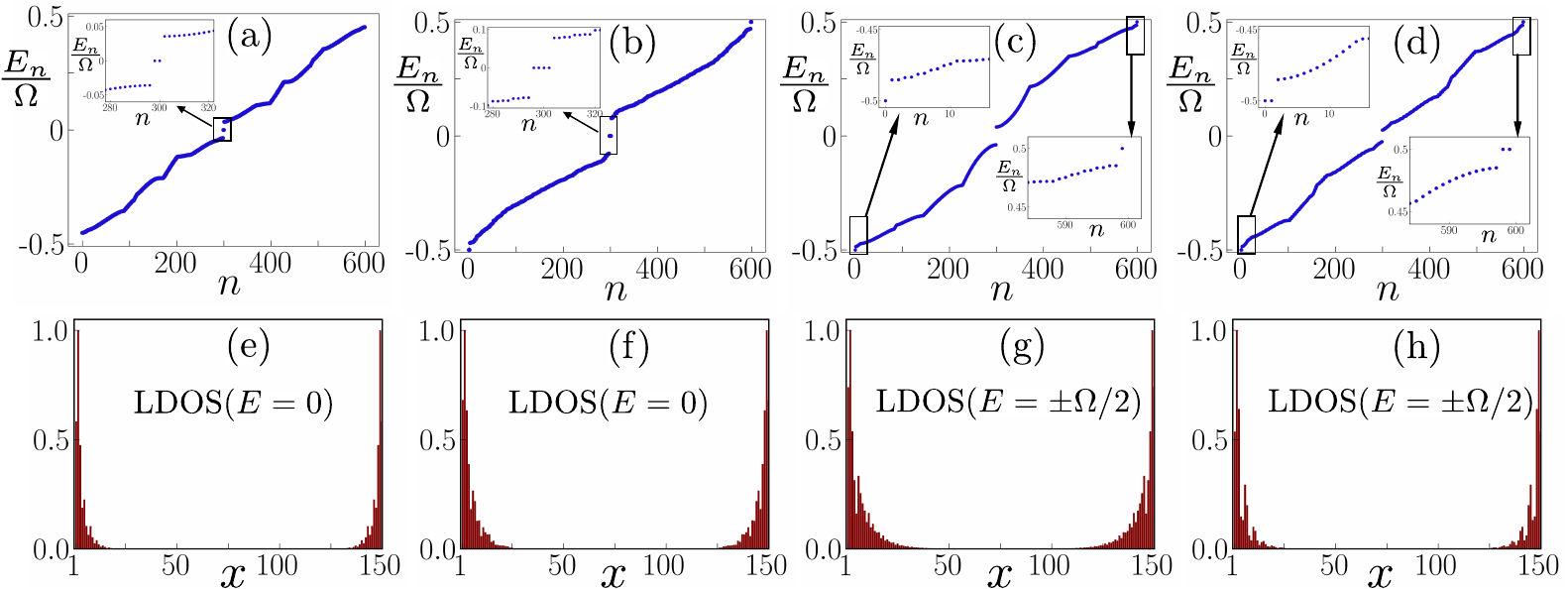}
	\caption{ \tbf {Quasi-energy spectra and LDOS for 0- and $\pi$-FMEMs:} In panels (a), (b), (c), and (d), we depict the behavior of real space energy spetra, $E_n$, as a function of state index, $n$, for $\mc{W}_0=1$, $\mc{W}_0=2$, $\mc{W}_\pi=1$, and $\mc{W}_\pi=2$, respectively. In panels (e) and (f), we illustrate the site dependent LDOS at $E=0$ corresponding to the panels (a) and (b), respectively, while in panels (g) and (h), we demonstrate the same at $E=\pm \Omega/2$ corresponding to the panels (c) and (d), respectively. Parameters of the model Hamiltonian (see Eq.~(1) and (2) of the main text), for the panels (a), (b), (c), and (d) are chosen as: $(V_0,\Omega)=(3,2.5)$, $(2.5,0.75)$, $(1,4.5)$, and $(4,1.75)$, while the rest of the parameters, $(\mu/t,\lambda_R/t, \Delta_0/t,J_A/t,t)=(1.5,0.5,0.3,0.4,1)$ remain same for all the panels.}
	\label{Fig.S1}
\end{figure}
\section{Analysis of Bulk gaps  in the driven system} \label{sec:Bulk_gap}
In the main text, we illustrate the behavior of DWNs, $\mc{W}_0$ and $\mc{W}_{\pi}$, for topological characterization of FMEMs in the $V \mhyphen \Omega$ plane with $\Omega$ and $V_0$ being the frequency and amplitude of the external drive. In a driven system, we can define two bulk gaps at quasienergies $E_\alpha=0$ as $G_0$, and $E_\alpha=\Omega/2$ as $G_\pi$ to corroborate with the $\mc{W}_0$ and $\mc{W}_{\pi}$, respectively. Here, we compute $G_0$ and $G_{\pi}$ in the $V\mhyphen \Omega$ plane and depict their behavior in Figs.~\ref{Fig.S2}(a)-(d) for the exactly same set of parameter values used to find the $\mc{W}_0$ and $\mc{W}_{\pi}$ as shown in Figs.~2(a)-(d) of the main text. Figs.~\ref{Fig.S2}(a),(b) represent the $G_0$ and $G_\pi$ when the Floquet drive is applied in the nontopological phase of the static Hamiltonian whereas, Figs.~\ref{Fig.S2}(c),(d) demonstrate the situation when static Hamiltonian hosts only AZMs. Interestingly, in Fig.~\ref{Fig.S2}(c), we find multiple gap closing and reopening transitions as we change the frequency ($\Omega$) and amplitude ($V_0$) of the drive, however, the corresponding $\mc{W}_0$ remains zero for any value of $\Omega$ and $V_0$ (see Fig.~2(c) of the main text). Nevertheless, this is not unusual, since a topological phase transition is always associated with the bulk gap closing and reopening transition. On the contrary, a gap closing transition might not always necessarily imply a topological phase transition.
\begin{figure}
	\centering
	\includegraphics[scale=0.65]{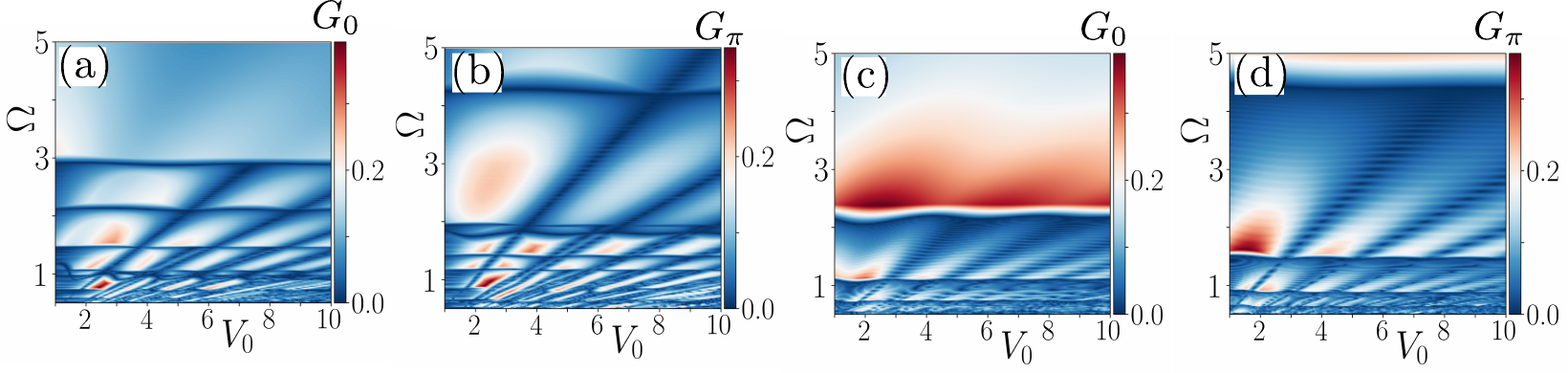}
	\caption{\tbf{Variation of bulk gaps at quasienergies $E=0$ ($G_0$) and $\Omega/2$ ($G_\pi$) in the driven system:} We depict the quasienergy gaps, $G_0$ (panels (a),(c)) and $G_\pi$ (panels (b),(d)), in the $V_0\mhyphen\Omega$ plane due to the application of time periodic drive. Panels (a) and (b)\,[(c) and (d)] correspond to the case when the static Hamiltonian is in the non-topological phase hosting no zero energy modes [only AZMs]. We choose $(\mu,J_A=1.5t,0.4t)$ in panels (a)-(b) and $(\mu,J_A=0,1.2t)$ in panels (c)-(d). We fix the other model parameters as $(t,\Delta_0,\lambda_R=1,0.3t, 0.5t)$ in all the panels which is same as mentioned in Fig.\,2 of the main text. }
	\label{Fig.S2}
\end{figure}

\section{Advantages of using altermagnets over external Zeeman field} \label{sec:AM_vs_ZeemanField}

\begin{figure}[h]
	\centering
	\includegraphics[scale=0.75]{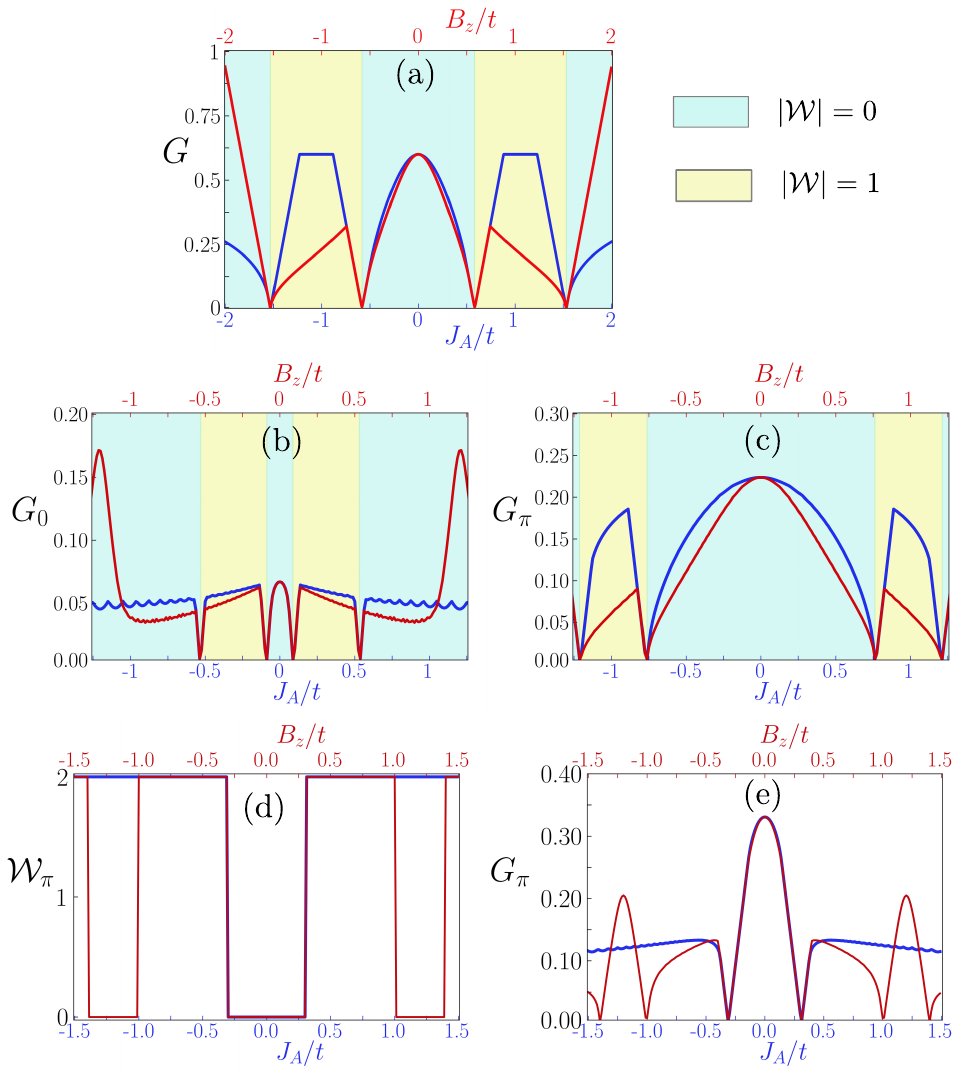}
	\caption{ \tbf{Comparison between AM and magnetic field using bulk gap and DWN, $\Wp$:} Panel (a): The variation of the bulk topological superconducting gap, $G$, in the static system is depicted as a function of strength of AM, $J_A/t$ (bottom axis) and strength of magnetic field (top axis). In the figure, blue and red lines indicate the variation of $G$ with $J_A/t$ and $B_z/t$, respectively. The light blue and yellow regions indicate the winding number with $|\mc{W}|=0$ and $|\mc{W}|=1$, respectively. Panels (b)-(c): The variation of bulk superconducting gaps, $G_0$ and $G_\pi$ in the driven system are shown as a function of $J_A/t$ (blue line) and $B_z/t$ (red line) with the $x$-axis convention similar to the panel (a). Panels (d)-(e): Dynamical winding number, $\mc{W}_\pi$ and associated bulk gap, $G_\pi$, are displayed as a function of $J_A/t$ (blue line) and $B_z/t$ (red line) with the $x$-axis convention similar to the panel (a). The model parameter values acros the panels are chosen as, panel (a): $(\mu/t,\lambda_R/t, \Delta_0/t,J_A/t,t)=(0.5,0.5,0.3,0.4,1)$, panels (b)-(c): $(\mu/t,\lambda_R/t, \Delta_0/t,J_A/t,t,V_0,\Omega)=(1.5,0.5,0.3,0.4,1,1.5,2.6)$, and panels (d)-(e): $(\mu/t,\lambda_R/t, \Delta_0/t,J_A/t,t,V_0,\Omega)=(0,0.5,0.3,0.4,1,1.5,2.6)$.}
	\label{Fig.S3}
\end{figure}
In the main text, we have mentioned briefly the advantages of using altermagnets (AMs) over external Zeeman field. In this section, we discuss these features in more detail.
\vskip 0.2cm
$\bullet$ \tbf{\underline{Model Hamiltonian for uniform magnetic field:}}  \vskip 0.2cm
$\ast$ \underline{Static case\,:} For the static case, we have considered the momentum space Hamiltonian in presence of a magnetic field, instead of AM, as,  
\begin{eqnarray}
	H_{\rm mag}(k) =& [t\cos (k) -\mu]\pi_z\sigma_0 + \lambda_R \sin (k) \pi_z \sigma_y 
	& + \,B_z \pi_0 \sigma_z + \Delta_0 \pi_x \sigma_0 \label{Eq.static_Ham_Bz}\ .
\end{eqnarray} 
The strength of the magnetic field is denoted by $B_z$, while definition of other model parameters remain same as mentioned in Eq.~(1) of the main text.

\vskip 0.1cm 
$\ast$ \underline{Driving protocol\,:} We periodically drive the above staic Hamiltonian (Eq.\,\eqref{Eq.static_Ham_Bz}) using the same time-periodic sinusoidal modulation to the chemical potential as it is in the case of AM, \ie  
$\mu(t) = \mu + V_0 \cos (\Omega t) $.

\vskip 0.1cm 
$\bullet$ \tbf{\underline{Advantages of using AM over magnetic field, $B_z$\,:}}
\vskip 0.25 cm
$\ast$ \underline{Larger topological superconducting gap:} \vskip 0.15cm  
Majorana modes in a topological superconductor are protected by the bulk superconducting gap. An enhanced gap provides additional stability against disorder and also increases the ambient temperature regime for the experimental observation of Majorana modes.  

First, in the static case, we compute the bulk gap $G$ for the AM using the Hamiltonian in Eq.\,(1) of the main text, and for the magnetic field using the Eq.\,\eqref{Eq.static_Ham_Bz}. In Fig.\,\ref{Fig.S3}(a), we depict the variation of $G$ as a function of  the strength of the AM, $J_A$ (bottom axis), and the magnetic field, $B_z$ (top axis). Notably, in the static case, the topological phase boundary remains identical for the AM and magnetic field, as indicated by the yellow (topological) and light-blue (non-topological) regions in Fig.\,\ref{Fig.S3}(a). Interestingly, within the topological phase (yellow region), the superconducting bulk gap $G$ for the AM case is consistently larger than that of the magnetic field case, and in some regions it becomes nearly twice as large in magnitude.  

Then, in the driven scenario, we compute the bulk gaps $G_0$ and $G_\pi$, associated with the 0-FMEMs and $\pi$-FMEMs, respectively, and display their variation with respect to $J_A$ in Figs.\,\ref{Fig.S3}(b) and (c). Here as well, we find that both $G_0$ and $G_\pi$ in the topological regions become always larger in the AM case than in the uniform magnetic field case, thereby providing enhanced stability in the driven system as well.

\vskip 0.2cm
$\ast$ \underline{Enhancement of topological boundary in the driven system:} \vskip 0.2cm
In addition to provide a larger topological bulk superconducting gap, the use of an AM also extends the topological phase boundary in the driven system compared to the case of using a magnetic field. We compute the invariant $\mc{W}_\pi$ [see Fig.\,\ref{Fig.S3}(d)] and the associated bulk gap $G_\pi$ [see Fig.\,\ref{Fig.S3}(e)], and compare their variation with the strength of the AM, $J_A/t$, and magnetic field, $B_z/t$. We find that for the AM the system is topological with $\mc{W}_\pi=2$ in the range $1 \le |J_A/t| \le 1.4$, whereas for the same strength of magnetic field the system remains non-topological with $\mc{W}_\pi=0$. The behavior of the bulk gap $G_\pi$ is fully consistent with this intricate behaviour.

Therefore, we believe that the use of an AM to realize Majorana modes offers significant advantages over a magnetic field or ferromagnet, both in static and driven case. For the sake of completeness, here we provide few parameter ranges for real materials.  From experimental point of view, $\rm InSb$ can serve as a possible candidate for the Rashba nanowire, leading to the following numerical estimates of model parameters ~\cite{Mourik2012Science}: $\displaystyle{ t (= \hbar^2 / 2m^* a^2)} \simeq 100  {\rm meV},\, \lambda_R \simeq 100  {\rm \mu eV}$, with a proximity-induced superconducting gap of about $250 {\rm \mu eV}$ at an external magnetic field of $0.15 , {\rm T}$, assuming $m^* \simeq 0.025 m_e$ and a lattice constant $a \sim 5 {\,\rm nm}$. On the other hand, ${\rm KV_2Se_2O}$~\cite{Jiang2025} or $\rm MnTe$~\cite{Lee2024PRL} may be possible to use as a probable candidate for altermagnet, exhibiting room-temperature altermagnetism with an energy scale in the range of $10$–$100{\,\rm meV}$. Although, we would like to stress that our model Hamiltonian and numerical simulations based on that 
do not correspond to the values of these real material parameters. Future experiments on AM materials and altermagnet-superconductor heterostructures can only settle such issue. 

\section{Stability of the driven topological superconductor (TSC) against disorder} \label{sec:DWN_disorder}

In the main text, we have shown the variation of DWNs, $\mc{W}_0$ and $\mc{W}_\pi$ in the $V_0\mhyphen \Omega$ plane considering the clean system (see Fig.~2 of the main text). Here, we expand this analysis further by incorporating onsite random static disorder in the driven system. We add to the clean time periodic Hamiltonian, $\mc{H}(t)=\mc{H}_0 + V(t)$ (see Eq.~(1) and (2) of the main text), a disordered term, $H_{\rm dis} = \sum_{x}\Psi_x^\dagger V(x) \pi_z \sigma_0 \Psi_x$ where $V(x)$ is random number generated from a box distribution $[-V_{\rm dis}/2, V_{\rm dis}/2]$ with $V_{\rm dis}$ being the strength of the disorder potential. We compute the DWNs in the presence of various disorder strength after performing average over 50 disorder configurations and present our results in Fig.~\ref{Fig.S4}(a)-(c). When we apply Floquet drive to the nontopological phase of the static model hosting no zero energy modes, the driven system anchors both $0$- and $\pi$- FMEMs.  For this case, the behaviour of $\mc{W}_0$ and $\mc{W}_\pi$ are displayed as a function of $\Omega$ in Fig.~\ref{Fig.S4}(a) and (b) for various stengths of disorder, $V_{\rm dis}$. Whereas,  Fig.~\ref{Fig.S4}(c) illustrates the same when the driven system hosts TAZMs, obtainded by periodically driving the trivial AZMs in the static mdoel. For the latter, as only $\pi$-FMEMs aquire topological nature, we only present the variation of $\mc{W}_{\pi}$. From these analysis, we find both $0$- and $\pi$-FMEMs sustain upto a sufficiently high disorder strength showcasing their topologically robust nature.
\begin{figure}[h]
	\centering
	\includegraphics[scale=0.6]{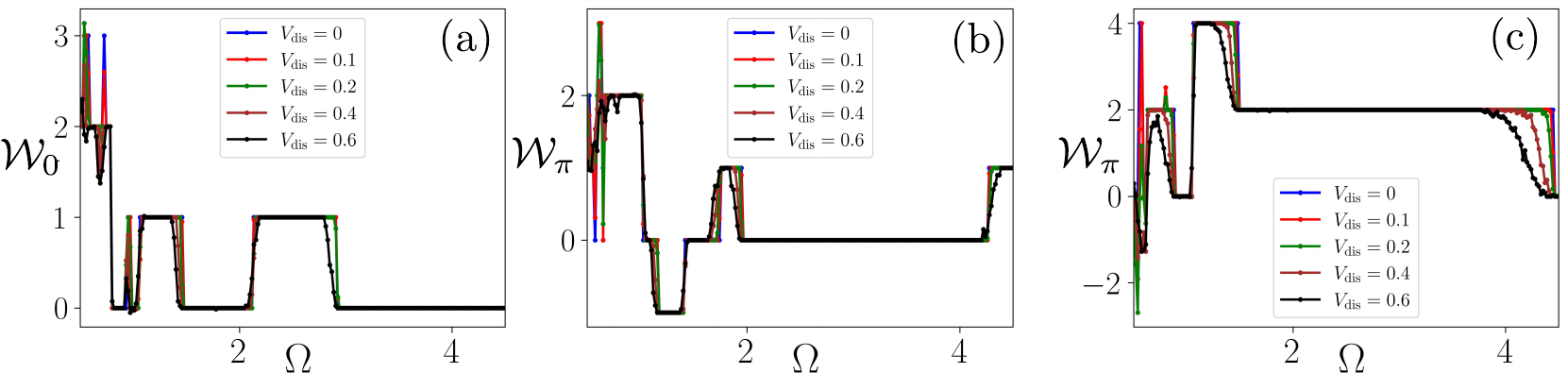}
	\caption{\tbf{Effect of static random disorder on the DWNs, $\mc{W}_0$ and $\mc{W}_\pi$:} In panels (a)-(c), we depict the variation of $\mc{W}_0$ and $\mc{W}_{\pi}$ as a function of $\Omega$ for various strengths of disorder, $V_{\rm dis}$. Panels (a) and (b) correspond to the case when Floquet drive is applied to the nontopological phase of the static model hosting no zero energy modes whereas in panel (c) we illustrate ${W}_{\pi}$ when the static model hosts only AZMs. In panels (a) and (b) we choose the other model parameters as: $(\mu,J_A=1.5t,0.4t)$, whereas in panel (c), $(\mu,J_A=0,1.2t)$. We fix the other model parameters as $V_0=2t,\Delta=0.3t,\lambda_R=0.5t$ across all the panels.}
	\label{Fig.S4}
\end{figure}

\section{Floquet Josephson current (JC) via numerical exact diagonalization}~\label{sec:Flq_JC_brute_force}
\begin{figure}
	\centering 
	\includegraphics[scale=0.43]{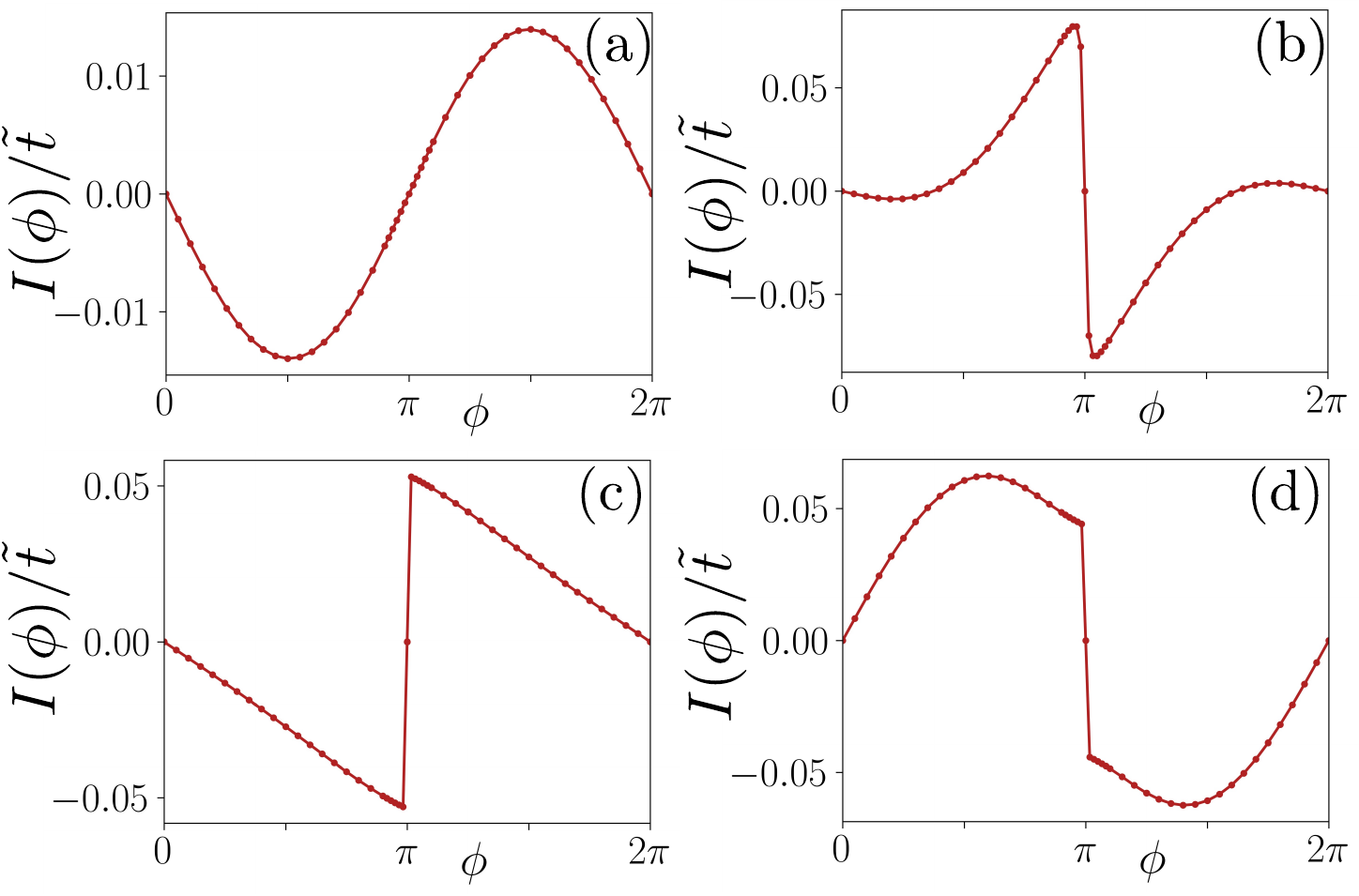}
	\caption{\tbf{JC in driven JJ employing numerical derivative of quasienergies:} In panels, (a)-(d), we present the variation of JC (obtained using Eq.~\eqref{Eq.JosephsonCurrent_Brute_Force}) as a function of $\phi$ considering four possible cases: (a) absence of FMEMs, (b) presence of only $0$-FMEMs, (c) presence of only $\pi$-FMEMs, and (d) presence of both $0$- and $\pi$-FMEMs. The model parameters $(\mu/t,J_A/t,V_0, \Omega)$ are chosen as:  (1.5,0.4,2,4) (panel (a)), (1.1,0.2,2,2) (panel (b)), (0.3,0.65,1.5,3)(panel c), and (1.1,0.2,3,1.5) (panel (d)). We fix the other model parameters as: $t=1,\lambda_R=0.5t,\Delta_0=0.3t,t_c=0.1t$ and $N_x=200$ lattice sites across all the panels.}
	\label{Fig.S5}
\end{figure}
In the main text, we have computed the JC in the driven JJ utilizing the energy resolved formalism~\cite{KunduPRL2024} (see Fig.~4 of the main text). This formalism provides a correct description of the occupation number of the FMEMs in the out-of-equilibrium scenario. In this section, we show that performing brute force derivatives of the quasienergy states with respect to the superconducting phase bias, $\phi$, may produce the discontinuous jump in the JC at $\phi=\pi$ signifying the $4\pi$-periodic Josephson effect but fails to identify the presence of $0$- and $\pi$-FMEMs separately. For this purpose, we use the following relation to obtain the JC in the driven system at zero temperature as, 
\begin{equation}
	I(\phi) = \sum_{E_\alpha<0} \frac{\partial E_\alpha(\phi)}{\partial \phi}\ , \label{Eq.JosephsonCurrent_Brute_Force}
\end{equation}
where, $E_\alpha$s are the quasienergies of the driven JJ obtained by diagonalizing the Floquet Hamiltonian, $\mc{H}_F$ (see Eq.~(3) of the main text). We compute the JC using the 
Eq.~\eqref{Eq.JosephsonCurrent_Brute_Force} and present our findings in Fig.~\ref{Fig.S5}. We consider four possible scenarios based on the presence or absence of $0$- and $\pi$-FMEMs, (i) absence of both $0$- and $\pi$-FMEMs (Fig.~\ref{Fig.S5}(a)), (ii) presence of only $0$-FMEMs (Fig.~\ref{Fig.S5}(b)), (iii) presence of only $\pi$-FMEMs (Fig.~\ref{Fig.S5}(c)), and (iv) presence of both $0$- and $\pi$- FMEMs (Fig.~\ref{Fig.S5}(d)). When either $0$- or $\pi$-FMEMs are present in the system, we find a discontinuity in JC at $\phi=\pi$, although we cannot distinguish between the topological $0$- and $\pi$-FMEMs which is only possible if one employs the energy resolved JC formalism~\cite{KunduPRL2024} as followed in the main text. When no FMEMs are present in the system, the JC exhibits usual $2\pi$-periodic behaviour due to the bulk states.

\section{Brief derivation of Floquet Josephson current formalism}~\label{sec:Flq_Josephson_Formalism}
In this section, we outline a brief derivation of the Floquet Josehson current formula. For a detailed derivation and understanding of the Floquet JC formalism, we strongly suggest the readers to look at the Ref.\,\cite{KunduPRL2024}. The key challenge is to find the occupation probabilities of the quasiparticle steady states as the external drive takes the system into the out-of-equilibrium phase. To begin with, we consider a periodically driven system which is weakly coupled to an external thermal reservoir with coupling strength, $V_\lambda$. Assuming the energy independent density of states of the resevoir \ie in the wide-band limit, the time dependent \Schr equation takes the following form,
\begin{equation}
	i\hbar \frac{d}{dt}\ket{\Psi(t)} = [\mc{H}(t) - i\Gamma^\lambda]\ket{\Psi(t)}\ . \label{Eq.Flq_Scrodinger_Eqn}
\end{equation}
Here, $\Gamma^\lambda=\pi V^\lambda\rho^\lambda {V^{\lambda}}^\dagger$ incorporates the density of states of the reservoir, $\rho^\lambda$, and the coupling between the reservoir and the system, $V^\lambda$. Owing to the periodicity of the driven system, \ie $\mc{H}(t+T) = \mc{H}(t)$ with $T$ being the time periodicity of the drive, we can write the solution of Eq.~\eqref{Eq.Flq_Scrodinger_Eqn} as, 
\begin{equation}
	\ket{\Psi_\alpha(t)} = e^{(-i E_\alpha/\hbar - \gamma_a)t} \ket{u_\alpha(t)}\ ,
\end{equation}
Here, $ \ket{u_\alpha(t)}$ is the time periodic Floquet modes and can be decomposed into Fourier modes as $ \ket{u_\alpha(t)} = \sum_{m\in \mathbb{Z}}  \ket{u_{\alpha}^{(m)}} e^{-im\Omega t}$ with $\Omega=2\pi/T$.

To this end, one can define the steady state density matrix, $\hat{\rho}(t)$ in the basis of Floquet modes, $ \ket{u_\alpha(t)}$ as, 
\begin{equation}
	\hat{\rho} = \frac{1}{T} \int_{0}^{T} dt \sum_{\alpha,\beta} \hat{n}_{\alpha\beta}(t)  \ket{u_\alpha(t)} \bra{u_\beta(t)}\ ,
\end{equation}
Then, the thermal average of $\hat{n}_{\alpha\beta}(t)$ can be expressed in terms of their Fourier modes as $n_{\alpha\beta}(t) =\sum_{q} e^{-iq\Omega t} n^{(q)}_{\alpha\beta}$ where the expression of $n^{(q)}_{\alpha\beta}$ is given by\,\cite{KunduPRL2024}, 
\begin{eqnarray}
	n^{(q)}_{\alpha\beta} = \sum_{ m}\int \frac{d\omega}{\pi} \frac{\bra{u_\alpha^{(m)}} \Gamma^\lambda \ket{u_\alpha^{(m+q)}} f^\lambda(\omega - \mu^\lambda)}{(\omega - E_\alpha^{(m)} + i \gamma_\alpha) (\omega - E_\beta^{(m+q)} -i \gamma_\beta)}\ ,
\end{eqnarray}
where, $ E_\alpha^{(m)} =  E_\alpha + m\Omega$ and $f^\lambda(x)= (1 + e^{x/\theta^\lambda})^{-1}$ 
is the Fermi distribution function with $\theta_r$ being the temperature of the reservoir. After a tedious algebra, one can show that in the weak coupling limit, \ie $\Gamma^\lambda\rightarrow 0$, $n^{(q)}_{\alpha\beta} = n_\alpha \delta_{\alpha\beta}\delta_{q0}$, where $n_\alpha$ is given by the simplified expression~ \cite{KunduPRL2024}, 
\begin{equation}
	n_{\alpha}(\mu_r) = \sum_m \braket{u_\alpha^{(m)}|u_\alpha^{(m)}} f^r(E_\alpha + m\Omega - \mu_r)\ .
\end{equation}

With this simplified expression for occupation of steady states, one can obtain the JC in the driven system averaged over the full time-period, $T$, as~\cite{KunduPRL2024}, 
\begin{equation}
	I_{0/\pi}(\phi) = \sum_\A n(E_\A,\mu^{0/\pi}_r) \frac{\partial E_\A(\phi)}{\partial \phi}\ , \label{Eq.FloquetJC}
\end{equation}
with $E_\alpha$ being the quasi-energy eigenvalues of the driven system. Importantly, the chemical potential of the reservoir plays the crucial role for probing $0$- and $\pi$-FMEMs of the driven system. 
One can probe and identify the presence of $0$- and $\pi$-FMEMs in the driven system distinctively by setting $\mu_r=0$ for $0$-FMEMs and $\mu_r=\Omega/2$ for $\pi$-FMEMs, thus providing the energy resolved formalism to identify the FMEMs~\cite{KunduPRL2024}.

\section{Floquet Josephson current for $\Wz=1$ and $\Wz=2$ } \label{sec:aaaaa}

\begin{figure}
	\includegraphics[scale=0.52]{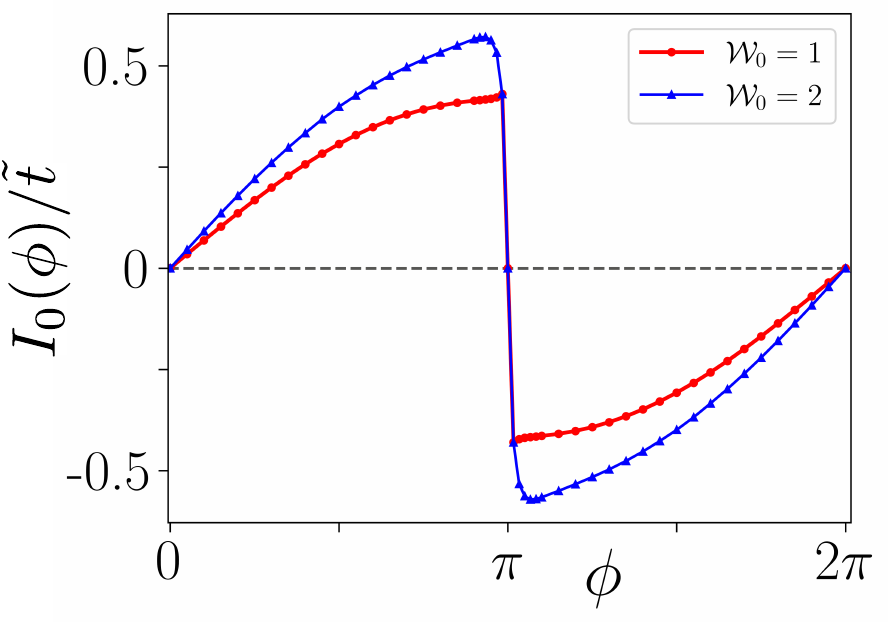}
	\caption{\tbf{Variation of JC with phase bias, $\phi$, for $\Wz=0$ and $\Wz=1$:} Variation of JC, $I_0(\phi)/\tilde{t}$ is depicted as a function of phase difference, $\phi$, for two topological phases, $\mc{W}_0=1$ and $\mc{W}_0=2$. 
		We set the model parameters as, $(\mu/t,\lambda_R/t, \Delta_0/t,J_A/t,\tilde{t}/t,t)=(1.5,0.5,0.3,0.4,0.01,1)$, choosing $(V_0,\Omega)=(3,2.5)$ for $\mc{W}_0=1$ and $(V_0,\Omega)=(2.5,0.75)$ for $\mc{W}_0=2$. } 
	\label{Fig.S6}
\end{figure}

In the main text, we have used the Josephson current (JC), both in the static and driven systems, to distinguish topological zero modes from non-topological zero-energy modes. 
For topological zero modes, the JC exhibits $4\pi$-periodicity as a function of the superconducting phase difference $\phi$, manifested as a discontinuous jump at $\phi=\pi$. On the other hand, for non-topological zero modes the JC remains $2\pi$-periodic in $\phi$. However, in principle, two topological phases with $\mc{W}=1$ and $\mc{W}=2$ cannot be distinguished by the JC, since in both cases the JC remains $4\pi$-periodic.  

Nevertheless, we have computed the Floquet JC (using Eq.\,(9) of the main text) in the driven system for the two topological phases $\mc{W}_0=1$ and $\mc{W}_0=2$, and the results are shown in Fig.\,\ref{Fig.S6}. We find that the magnitude of the discontinuous jump is larger for $\mc{W}_0=2$ compared to $\mc{W}_0=1$. This arises because in the $\mc{W}_0=1$ phase only the two zero modes, localized at the junction (see Fig.\,1 of the main text), contribute to the JC, whereas in the $\mc{W}_0=2$ phase four zero-energy modes are localized at the junction, thereby leading to higher contribution to the JC. However, this feature is not at all universal and heavily dependent on the choice of parameter regime. Therefore, drawing any conclusion from the height of the discontinous jump is not justified. We would like to further emphasize that this is only a comparative feature between the $\mc{W}_0=1$ and $\mc{W}_0=2$ phases and JC signature cannot be used as a definitive criterion to distinguish between them.  One possible way to distinguish between the $\mc{W}_0=1$ and $\mc{W}_0=2$ phases can be the zero-bias differential conductance [$dI/dV(eV=0)$]~\cite{Mondal2024}.


\begin{figure}
\centering
\includegraphics[scale=0.6]{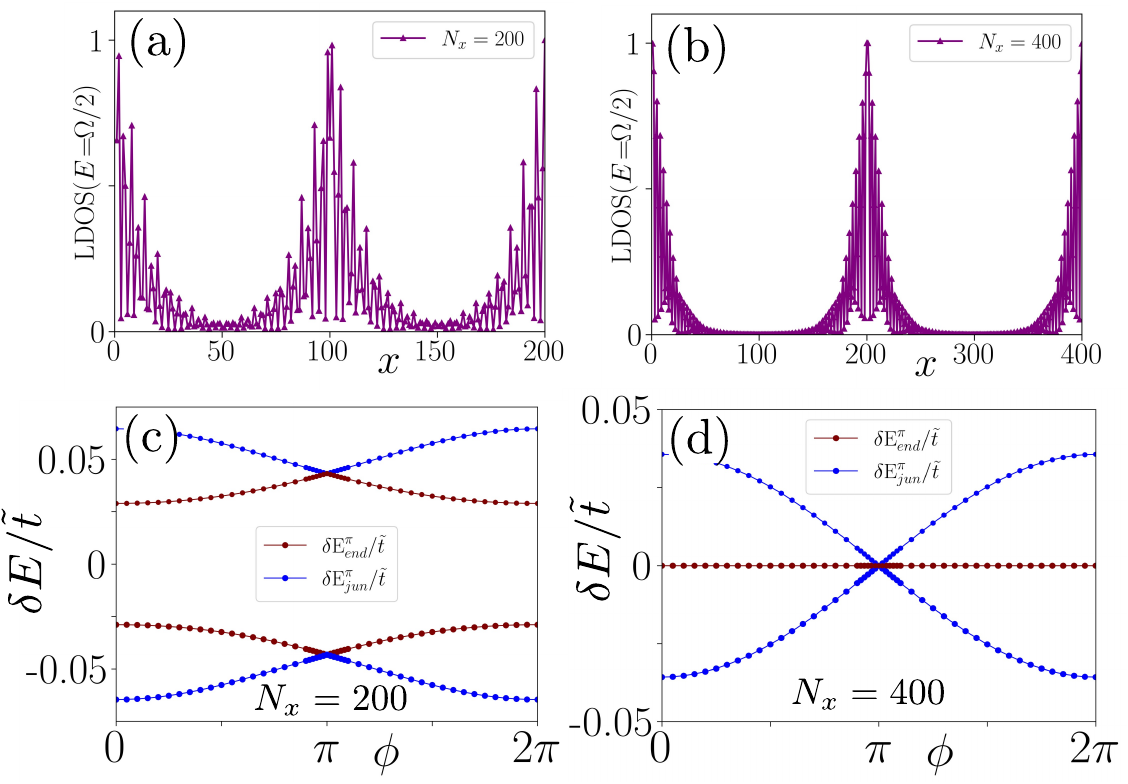}
\caption{\tbf{LDOS and $E\mhyphen \phi$ relation due to the TAZMs in the driven JJ:} In panels (a) and (b), we depict the LDOS as a function of system coordinates, $x$, corresponding to TAZMs maintaining the phase bias at $\phi=\pi$. In panels (c) and (d), we present the energy deviation of TAZMs from $E=\Omega/2$ for both near ($\delta E^\pi_{jun}/\tilde{t}$) and far away ($\delta E^\pi_{end}/\tilde{t}$) localized modes from the junction. In panels (a) and (c), we fix $N_x=200$ lattice sites, whereas in panels (b) and (d) $N_x=400$ lattice sites where $N_x$ is the length of the JJ. We choose the other model parameters across all the panels as $(\mu,J_A,\Delta_0,\lambda_R,V_0,\Omega=0,1.2t,0.3t,0.5t,3.1t,3)$. }
\label{Fig.S7}
\end{figure}

\vspace{-0.5cm}

\section{Local density of states (LDOS) and energy deviation of topological accidental 
zero modes (TAZMs) in driven Josephson junction (JJ)} \label{sec:Ldos_JJ}
%

In the main text, we have computed the $4\pi$-periodic JC for FMEMs (see Fig.~4 of the main text). 
We find FMEMs exhibit $4\pi$-periodic JC signature only when the length of the JJ is larger than a critical length, $N_x^c$. On the other hand, for TAZMs, which are $\pi$-FMEMs obtained by periodically driving the trivial AZMs of the static Hamiltonian~\cite{Mondal2025PRBL}, $N_x^c =300$ lattice sites, whereas for the conventional FMEMs $N_x^c=100$ lattice sites. The larger value of $N_x^c$ for TAZMs can be attributed to the longer Majorana localization length, $\xi_m$. In this section, we explicitly demonstrate such behavior by computing the local density of states (LDOS) for the TAZMs in the JJ setup, maintaining the superconducting phase difference, $\phi$, fixed at $\pi$. We illustrate the LDOS distribution as a function of system coordinates ($x$) in Figs.\,\ref{Fig.S7}(a) and (b) for two system sizes $N_x=200<N_x^c$ and $N_x=400>N_x^c$, respectively. We find that when $N_x<N_x^c$, the TAZMs localized away from and near the junction, overlap with each other leading to the usual $2\pi$-periodic Josephson effect (see Fig.4(c) of the main text). In sharp contrast, when $N_x=400>N_x^c$, the TAZMs display the $4\pi$-periodic Josephson effect as the far away modes do not overlap with the modes localized near the junction. To further illustrate this point, we compute the energy deviation of TAZMs from $E=\Omega/2$ in JJ setup as a function of phase bias, $\phi$. Due to presence of weak coupling, $\tilde{t}$ between the two TSCs forming the JJ, energy of the TAZMs deviates from $E=\Omega/2$. For $N_x>N_x^c$, the FMEMs, localized near the junction, become degenerate at $\phi=\pi$ and undergoes a parity change at $\phi=\pi$ resulting in $4\pi$-perodic JC. in contrast, when $N_x<N_x^c$, TAZMs which are broadly localized near and far away from the junction, fuse among themselves and cannot become degenerate at $\phi=\pi$. This intricate behavior is presented in Figs.~\ref{Fig.S7}(c) and (d). 	

\end{onecolumngrid}	

\end{document}